\documentclass[12pt]{ar}
\usepackage[english]{babel}
\usepackage{psfig,graphicx}
\usepackage{supertab}
\onecolumn

\begin{document}

\title{Spectral variability of the peculiar A-type supergiant 3\,Pup}

\author{Eugenij Chentsov$^1$, Valentina Klochkova$^1$ \& Anatoly Miroshnichenko$^2$}

\institute{1 -- Special Astrophysical Observatory RAS, Nizhnij Arkhyz, 369167  Russia \\
          2 -- University of North Carolina at Greensboro, NC       
         27042, USA}

\date{\today}	     

\abstract{Optical spectra of the peculiar supergiant 3\,Pup taken in
1997--2008 are used to study the spectral peculiarities and velocity field
in its atmosphere. The profiles of strong FeII lines and of the lines of
other ions have a specific shape: the wings are raised by emissions,
whereas the core is sharpened by a depression. The latter feature becomes
more pronounced with the increasing line strength, and the increasing
wavelength. Line profiles are variable: the magnitude and sign of the
absorption asymmetry, and the blue--to--red emission intensity ratios vary
from one spectrum to another. The temporal Vr variations are minimal for
the forbidden emissions and sharp shell cores of the absorption features
of FeII(42), and other strong lines of iron-group ions. Therefore the
average velocity for the above lines can be adopted as the systemic
velocity: V$_{\rm sys} = 28.5 \pm 0.5$\,km/s. The weakest photospheric
absorptions and photospheric MgII, SiII absorptions exhibit well-defined
day-to-day velocity variations of up to 7\,km/s. Quantitative spectral
classification yields the spectral type of A2.7\,$\pm$\,0.3~Ib. The
equivalent widths and profiles of H$\delta$ and H$\gamma$, and the
equivalent width of the OI~7774\,\AA{} triplet yield an absolute magnitude
estimate of M$_{\rm v}=-5\lefteqn{.}^m5 \pm 0\lefteqn{.}^m3 $,
implying the heliocentric distance of 0.7\,kpc. }

\authorrunning{Chentsov, Klochkova  \&  Miroshnichenko}
\titlerunning{Spectral variability of 3\,Pup}

\maketitle

\section{Introduction}

The nature of the bright star 3\,Pup
(HD\,62623\,=\,HR\,2996\,=\,HIP\,37677) has long remained unclear. In the
Bright Star Catalog [1] this star is listed as a A2\,Iabe--supergiant.
However, 3\,Pup was also viewed as an object related to stars evolving
towards planetary nebulae with binary core~[2]. The spectroscopic study of
Plets~et~al.~[3], who found the star's atmosphere to have a close-to-solar
chemical composition, provided new evidence for the massive supergiant
nature of 3\,Pup. However, the star has a property that is
uncharacteristic to massive A-type supergiants: a circumstellar dust
shell. The shell shows up primarily in the form of IR--excess (3\,Pup was
identified with the IR--source IRAS\,07418$-$2850) and specific two-peaked
emission features in the optical spectrum.

The star's magnitudes are B\,=\,4$\lefteqn{.}^{\rm m}12$ and
V\,=\,3$\lefteqn{.}^{\rm m}98$, and the color excess
E(B--V)\,=\,0$\lefteqn{.}^{\rm m}08$~[4]. Its Galactic and equatorial
J2000 coordinates are l\,=\,244$\lefteqn{.}{\degr }4$,
b\,=\,--2$\lefteqn{.}{\degr}5$ and $\alpha$\,=\,$07^{\rm h} 43^{\rm m}
48^{\rm s}$, $\delta =-28{\degr}$57${'}$\,17${''}$, respectively. The recently
refined parallax is $\pi$\,=\,0.59$\pm$0.17\,mas~[5]. Because of its
high brightness 3\,Pup was chosen as one of the several hundred stars used
as the basis of Harvard spectral classification. However, whereas
Cannon~[6] assigned to it the same spectral type A2\,pec as to
$\alpha$\,Cyg (the peculiarity of both stars showed up in the form of too
narrow and weak hydrogen lines, i.e., both stars are supergiants),
Maury~[7] noticed that 3\,Pup has a somewhat lower temperature and
luminosity than $\alpha$\,Cyg, A2--A3 and Ib--II, respectively, in the
currently adopted notation.

Merrill~[8] found  some emission lines in the spectrum of 3\,Pup: a
double H$\alpha$ line with a relatively stronger red component (a P Cyg
III type profile according to Beals~[9]), [OI]\,1F\,6300, 6364\,\AA{},
FeII(40)\,6433, 6516\,\AA. He also suspected the presence of emission
features in some other FeII lines. Emission features in the profiles of
H$\beta$, H$\gamma$~[10] and NaI(1)~doublet~[3] were noticed only 60 years
later. The known infrared emission features include the strong
CaII\,(2)\,8498, 8542, 8562\,\AA{} triplet~[11], and the [CaII]\,1F 7291,
7324\,\AA{} doublet~[12] (these lines were erroneously attributed to
[OII]). As for absorption lines, the eye estimates of their intensities in
the 3120--6565\,\AA{} wavelength interval~[13, 14] confirmed a lower
excitation temperature of 3\,Pup, compared to that of $\alpha$Cyg, as it
followed from classification of Maury, and revealed shell signatures in
the star's spectrum. In particular, the MgII\,(4) and SiII\,(1,2,3)
absorptions in the spectrum of 3\,Pup are weaker than the same absorptions
in the spectrum of $\alpha$\,Cyg, whereas the FeII\,(42) lines are enhanced
compared to FeII\,(37, 38, 49) lines. The differences between the spectra
of a supergiant and a shell star are readily seen in Plate~22 in the atlas
of Morgan et al.~[15], where they are illustrated on the examples of
$\alpha$\,Cyg~(A2~Ia) and 17\,Lep (Ap sh). The temperature of 3\,Pup is
close to the temperatures of both stars, but its spectrum is closer to
that of 17\,Lep by a number of features (the depth of TiII and FeII lines
and the widths of hydrogen lines).

It became clear from the first inspection of the available spectra of
3\,Pup that it must be a very promising object for high-resolution
spectroscopy: the star is bright and its spectrum abounds in various,
mostly weak, but well-defined features. But however strange it may appear,
its optical and near infrared spectrum has not yet been thoroughly
described. This circumstance led us to include 3\,Pup into our list of
program stars with circumstellar shells for a detailed spectroscopic
analysis in order to study the expected spectral variability.

\section{Observations and reduction of spectra}

We obtained the spectra of 3\,Pup with the NES echelle spectrograph mounted
at the Nasmyth focus of the 6-meter telescope of the Special Astrophysical
Observatory~[16, 17]. Observations were made with a 2048$\times$2048~CCD
and an image slicer~[17]. The spectroscopic resolution and the
signal-to-noise ratio are $\lambda / \Delta\lambda$\,=\,60000 and
S/N$>$100, respectively. To extend the time interval of observations of
3\,Pup, we use the intermediate-resolution spectra that we took in
1997--1998 in the prime focus of the 6-m telescope with the PFES echelle
spectrograph~[18]. We used a modified ECHELLE context~[19] of the MIDAS
package to extract one-dimensional vectors from the two-dimensional
echelle spectra. Cosmicray hits were removed via median averaging of two
successively taken spectra. Wavelength calibration was performed using the
spectra of a hollow-cathode Th-Ar lamp.

In addition to the spectra taken with the NES spectrograph, we used
high-resolution spectra of the star taken with the ESPaDOnS
spectrograph~[20] of the 4-m CFHT telescope, and the cs23
spectrograph~[21] operating in the coude focus of the 2.7\,m telescope of
the McDonald observatory. The latter spectra were reduced using the IRAF
software package. Table~1 lists the dates of all observations, recorded
regions of the spectrum, the spectrographs employed, and the spectral
resolution. Table~2 lists the inferred heliocentric radial velocities Vr.

We used the telluric [OI], O2 , and H$_2$O lines to control and correct the
instrumental matching of the spectra of the star and the hollow-cathode
lamp. The last column of Table~1 indicates whether a correction has been
applied, and the residual errors for each spectrum. For a more detailed
description of the procedure of radial velocity Vr measurement from the
spectra taken with the NES spectrograph, and the sources of errors,
see~[22]. The r.m.s. error of the measured Vr for stars with narrow
absorption lines in the spectrum is less than 0.8\,km/s (the accuracy of
the velocity inferred from a single line).

\section{Discussion of the results}

\subsection{Peculiarity of the optical spectrum of 3\,Pup}

A comparison of the line profiles and parameters in our spectra of 3\,Pup,
and in the spectra of the comparison stars obtained with the same spectral
resolution showed that the shell appears not solely in the form of several
lines mentioned above. The shell signature grows gradually both with the
line intensity and line wavelength. Figure~1 compares the central depths
of absorption lines in the spectra of 3\,Pup and $\alpha$\,Cyg (the depth
corresponding to zero residual intensity is assumed to be equal to 100).
Whereas MgII and SiII lines at any intensities and in any part of the
optical spectrum are deeper for $\alpha$\,Cyg than for 3\,Pup, the TiII
and FeII lines of low and intermediate intensity located in the blue part
of the spectrum are deeper for $\alpha$\,Cyg, but with the increasing
intensity, the depths of these lines in the spectra of the two stars
become closer to each other and the strongest TiII(13) and FeII(42) lines
in the spectrum of 3\,Pup are deeper than the corresponding lines in the
spectrum of $\alpha$\,Cyg. In the yellow part of the spectrum of 3\,Pup,
FeII absorptions are anomalously weak (the bottom left corner of Fig.\,1),
they are all to a certain extent ``raised'' by emissions, in some of them
the absorption core of the profile is uplifted to the continuum level.
Note that in this case the MgII and SiII absorptions in the spectrum of
3\,Pup are weaker than the corresponding absorptions in the spectrum of
$\alpha$\,Cyg both in terms of depths and equivalent widths (which in the
case of MgII\,4481 are equal to 0.56 and 0.66\,\AA{}, respectively), and
the equivalent widths of the TiII\,(13) doublet are approximately the same
in the spectra of both stars. As it is evident from Fig.\,2, the smaller
widths of these lines in the spectra of 3\,Pup are compensated by their
greater depth. The profiles of the TiII\,(13) and FeII\,(42) lines have
the same shape. In Fig.\,3 we compare the profiles of FeII lines of
different intensity in the spectra of 3\,Pup and $\alpha$\,Cyg and with the
profiles of the MgII\,4481 line. In $\alpha$\,Cyg all lines form near the
photosphere, they are symmetric and differ only in depth. The only true
photospheric line in the spectrum of 3\,Pup is MgII 4481 A, whereas FeII
features exhibit evident contribution of the shell, which gives the
profiles their specific shape: the wings are raised by emissions, whereas
the line core is sharpened. The latter shows up even for absorptions with
depths R\,$\approx$\,20 and becomes stronger with increasing line
intensity, thereby explaining the upward bends of the symbol chains in
Fig.\,1. Pointed cores are also typical for H$\beta$--H8 Balmer lines,
their central depths are smaller than the corresponding central depths in
the spectra of $\alpha$\,Cyg.

Emissions in the wings become apparent in high resolution spectra starting
at least from $\lambda \approx$\,4300 A. Figure~4 illustrates the
variation of the line profile with wavelength: as we move from the blue
part of the spectrum towards its red part, the absorptions gradually give
way to the two-peaked emissions with an intensity gap at the center. The
profile of the double emission of FeII(40)\,6433\,\AA{} (the upper profile
in Fig.\,4) repeats in Fig.\,5, as compared to the profiles of the
CaII(2)\,8542\,\AA{} permitted line and several forbidden lines. In the
spectrum of 3\,Pup the latter are represented not only by the [OI]\,1F and
[CaII]\,1F doublets mentioned above, but also by numerous weak [FeII]
lines. A comparison of the emission profiles provides important
information about the geometry and kinematics of the 3\,Pup shell. All
forbidden emissions in the accessible region of the spectrum have equally
shaped profiles with the same widths, and yield the same radial velocity
within the errors. Permitted emissions differ from forbidden emissions by
greater intensity gradients in the central parts of the profiles, they are
appreciably broader (for [OI]\,6300 and FeII\,6433\,\AA{} the full widths
at the continuum level are equal to 120 and 140\,km/s, respectively) and
are often systematically offset with respect to forbidden emissions (by
4\,km/s on December, 25, 2004). The strong forbidden CaII\,(2) triplet has
rather interesting two-step profiles (in Fig.\,5 it is represented by the
emission at 8542 A): the narrow double peak is appreciably shifted redward
relative to the broader ``pedestal''. The radial velocity of the peak is
close to the velocities of forbidden emissions, whereas that of the lower
component of the profile, to the velocities inferred from permitted
emissions.

Line profiles vary with time. The most evident spectrum-to-spectrum
variations are those of the magnitude and sign of the asymmetry of
absorptions and the blue-to-red emission intensity ratios. Figure~6 shows
these variations for the case of FeII lines and the D1\,NaI\,(1) line. The
latter is less blended with atmospheric water absorptions than D2 line,
and its interstellar component shows a better outlined two--component core.
The stellar and interstellar components are easy to separate because of
the variability of the former. The differences between the profiles
disappear at the residual intensity of r\,$\approx$\,85. Above this level we
observe the stellar NaI line, the shape of its profile and its
evolution with time resemble what we see for the FeII\,5363\,\AA{} line;
below this level we observe the interstellar line, which, unlike what we
see in the case of FeII\,(42) lines, has almost vertical and stable slopes,
they merge in the 85$> \rm r >$15 interval.

\subsection{Radial velocity variations}

The anomalies of line profiles in the spectrum of 3\,Pup affect the
inferred radial velocities, which depend on the features and methods used
to measure them, thereby making it difficult to compare our data with
the published results of other authors.

The Lick spectrograms of 1908--24 used by Johnson and Neubauer~[23] to
classify 3\,Pup as a small amplitude spectroscopic binary were obtained on
a blue-sensitive emulsion with an average dispersion of about 20\,\AA{}/mm
and were measured visually. The inferred velocities appear to refer to the
region of the profiles of relatively strong lines with maximal
photographic density gradients. The proposed orbital elements
(P\,=\,137$\lefteqn{.}^{\rm d}$767, K\,=\,3.6\,km/s,
$\gamma$\,=\,25.4\,km/s) have been reproduced for more than 60~years in
the catalogs of spectroscopic binaries including the most recent such
catalog~[24]. Swings~[13] used panchromatic plates and found no velocity
variations with time or from one group of lines to another. He inferred a
mean radial velocity of Vr\,$\approx$\,31 km/s. According to Crespin and
Swensson~[14], Vr\,$\approx$\,27\,km/s.

In addition, several radial velocity of 3\,Pup have been published that
were measured from high-resolution (up to 100000) spectra, but,
unfortunately, the authors did not mention the measurement method.
Loden~[25] used CCD echelle spectra taken during five successive nights in
February, 1991. He used 80 lines in the wavelength interval
3850--4870\,\AA{} to infer an average radial velocity of
Vr\,=\,28.7$\pm$0.2\,km/s (in their more recent paper Loden and
Sundman~[26] corrected slightly this result: 29.3$\pm$1.0\,km/s).

Plets et al.~[3] also had five CCD spectra at their disposal, which were
taken from 1986 through 1991 in small, 30--50\,\AA{} wide, spectral regions
between 3920 and 6180\,\AA{} with 5 to 10 lines in each region. The data
for four of the five dates yield velocities that range from 27.7 to
30.9\,km/s, i.e., are close to the result obtained by Loden~[25]. However,
the above authors draw different conclusions. Loden~[25] found neither the
variation of velocity with time nor differential line shifts beyond the
measurement errors. Plets et al.~[3] combined the new Vr value with some
of Johnson and Neubauer data~[23] to corroborate the spectroscopic
binarity of 3\,Pup.

We performed our radial velocity measurements via a mutual wavelength
shift so as to align the direct and reversed images of the line profile.
In this way, we can determine the velocities of individual features of the
line profile. The list of lines used to measure the velocities can be
found in Table~3. The complete electronic version of this table is
available at www.sao.ru/hq/ssl/3Pup-lines.ps and
www.sao.ru/hq/ssl/3Pup/Table3.html. It contains the results of
identification of the features; the adopted laboratory wavelengths;
residual intensities ``r'' (central values for the absorptions and peak
values for the emissions), and the heliocentric radial velocities for the
absorption cores or for the whole emissions. Residual intensities r are
rounded here to hundredths (0.01) and radial velocities Vr, to 1 km/s. The
table contains only the data based on high-resolution spectra with the
velocities controlled and corrected by telluric lines. The residual
discrepancy of the zero points of the radial velocity scales for
different observing dates does not exceed 0.6\,km/s if estimated from
interstellar NaI\,(1) lines.

The top rows of Table~2 give the Vr values for NaI and other interstellar
lines and bands. The uppermost row lists the velocities of the
interstellar components of NaI\,(1) lines as a whole (i.e., averaged over
all our measurements: 33.3$\pm$0.2\,km/s). Below we list the velocities of
the two components (one under an other) in the cases where both can be
discerned. The short-wavelength component with Vr\,$\approx$\,28\,km/s is
stronger in KI\,(1) and CaII\,(1) lines and the longwave component with
Vr\,$\approx$\,37\,km/s, in NaI\,(1) lines. Note that the velocities of diffuse
interstellar bands (DIB) are determined from different sets of these
bands because of the different operating wavelengths intervals.

Does the absorption depth correlate with the velocity measured from the
same absorption? Figure~7 shows the Vr(r) curves based on the data from
Table~3. The strongest systematic trend of velocity with depth was
observed on December, 25, 2004. For this date the positions of individual
lines are shown on the plot: the weakest FeII, TiII, CrII absorptions
(bottom left), the FeII\,(42) triplet (top right), and MgII, SiII lines
(below the iron-group lines). The vertical scatter of symbols gives an
idea of the accuracy of velocity measurements based on lines of different
intensity. For the dates with appreciably smaller velocity differences
than on December, 25, 2004 we show in Fig.\,7 only the generalized Vr(r)
curves. The dependencies for 2006 and 2007, which are very close to each
other, are shown by a single, almost horizontal line. We used the Vr(r)
curves in the generalization and averaging the radial velocity data and
list the results in Table~2.

Our measurements suggest that the most stable feature is the outer shell
of 3\,Pup. Velocity variations with time are minimal, and the velocity
values for forbidden emissions and sharp shell cores of FeII\,(42)
absorptions and other strong iron-group ion lines are close to each other.
The mean velocity averaged over the above lines (rows 9, 10 and 12--14 in
Table 2, respectively) can be adopted as the systemic radial velocity for
the system as a whole: V$_{\rm sys}$\,=\,28.5$\pm 0.5$\,km/s, which differs
from both likely values of $\gamma$\,=\,25.6 and 23.5\,km/s implied by the
model proposed by Plets et al.~[3].

On the other hand, the weakest photospheric absorptions
(r\,$\rightarrow$\,1, row 16 in Table~2) show well defined date-to-date
velocity variations amounting to 7\,km/s between December, 25, 2004 and
December, 26, 2006. As is evident from a comparison of rows 15 and 16 of
Table~2, stronger and photospheric MgII and SiII absorptions show similar
velocities and similar velocity variations with time. The latter fact is
immediately apparent even for permitted lines (row 11 in Table 2), most of
which are slightly blueshifted relative to weakening absorptions. These
facts, like the narrowing of forbidden emissions compared to permitted
emissions, are consistent with the interpretation of 3\,Pup as a star with
an equatorial disk and a low-mass companion~[3]. Our Vr values for
photospheric absorptions lie within the limits imposed by the radial
velocity curve based on the data of Johnson and Neubauer~[23], but our
fragmentary observations are insuficient for its validation and
correction.

The velocities of the absorption cores of the first Balmer series lines,
except H$\alpha$, appear to be close to the velocities of strong FeII
absorptions, i.e., to V$_{\rm sys}$. The spectra taken on December, 25,
2004 and December, 26, 2006 show a small Balmer progress. More conclusive
evidence for the radial gradient of velocity in the shell of 3\,Pup is
provided by the H$\beta$ (November, 4, 2008) and H$\gamma$ (December, 25,
2004) profiles. These profiles exhibit not only the main cores used
to infer the velocities Vr listed in Table~2, but also the blueshifted
components with Vr\,$\approx$\,0\,km/s (see Fig.\,8). The extra-atmospheric
ultraviolet resonance lines are shifted with respect to the photospheric
lines even more, by ($-50 \div -60$)\,km/s~[10].

\subsection{Spectral type and distance}

The blue part of the spectrum of 3\,Pup, where emission components of the
lines are weak, resembles the corresponding regions in the spectra of
A2--3\,Ib-type stars, and, in particular, those of HD\,207673 and
BD\,+60$^{\degr}$2542~(A2\,Ib), HD\,210221~(A3\,Ib), and
HD\,13476~(A3\,Iab) as reported by Verdugo et al.~[27, 28], who observed
it with a resolution which is close to that of our spectra. Given that the
profiles of metal lines suffer from distortions that increase with
wavelength and line intensity, we performed quantitative classification by
shallow FeI, FeII, and TiII absorptions and by the MgI and MgII lines in
the 4280--4700\,\AA{} wavelength interval. The calibration relations of
the equivalent width as a function of spectral type W(Sp) are based on the
equivalent widths of A0--F0-type supergiants from~[29]. In addition, we
also compared the ratios of the central depths in the blue parts of the
spectra of 3\,Pup and $\eta$\,Leo~(A0\,Ib), HD\,21389~(A0\,Ia),
$\alpha$\,Cyg~(A2\,Ia), HD\,17378~(A5\,Ia) determined using NES
spectrograph. We estimated the luminosity class and Mv from the equivalent
widths and profiles of H$\beta$ and H$\gamma$ (we show the latter for
3\,Pup and comparison stars in Fig.\,9). We also estimated the luminosity
from the equivalent width of the OI\,7774\,\AA{} triplet (W\,=\,1.42\,\AA{},
which corresponds to luminosity class Ib). Our final estimates of the
spectral type and absolute magnitude of 3\,Pup are A2.7$\pm$0.3\,Ib and
Mv\,=\,$-5.5 \pm0.3^{\rm m}$, respectively.

The trigonometric parallax of 3\,Pup, which corresponds to d$>$1.4\,kpc,
is too small (about 0.6\,mas) and therefore too uncertain. The distance
estimates based on the Av(d) dependencies derived by Neckel and Klare~[30]
and Plets et al.~[3] are equally unreliable. They indicate that
interstellar extinction increases slowly with distance towards 3\,Pup, and
that it reaches Av(3\,Pup)$\approx 0.5^{\rm m}$ at d\,$\approx$\,1.4\,kpc,
whereas both the stellar and nonstellar radial velocities of 3\,Pup are
observed already at 0.3--0.5\,kpc. Currently, the most accurate results
for 3\,Pup are provided by the method of spectroscopic parallax, which
yields d\,$\approx$\,0.7\,kpc for the E(B--V) and Mv values mentioned
above.

\section{Conclusions}

In the case of FeII lines in the spectrum of 3\,Pup there is an evident
contribution of the shell, which gives the profiles their specific shapes:
the wings are raised by emissions, whereas the core is sharpened. The
latter is apparent even for the absorptions with depths R\,$\approx$\,20
at $\lambda>4300$\AA{} and becomes more pronounced with increasing line
strength. Sharpened cores are typical of Balmer lines except for H$\alpha$.

The profiles of the CaII\,(2) triplet have a ``twostep'' shape: a
narrow double peak is appreciably redshifted with respect to the relatively
broader ``pedestal''. The radial velocity of the peak is close to the
velocities of forbidden emission lines, whereas that of the lower
component of the profile is close to the velocities determined from the
permitted emissions.

Line profiles vary with time: both the magnitude and the sign of the
absorption asymmetry, and the blue--to--red emission intensity ratios vary
from one spectrum to another. The profiles of all forbidden emissions
in the recorded region of the spectrum have the same shapes and widths and
yield the same radial velocity within the errors.

Radial velocity variations of forbidden emissions and sharp envelope cores
of FeII\,(42) absorptions and other strong iron-group ion lines are minimal
and the velocities of these lines are close to each other. The mean
velocity averaged over the above lines can be adopted as the radial
velocity of the system as a whole: V$_{\rm sys}$\,=\,28.5$\pm$0.5\,km/s.

The weakest photospheric absorptions (with residual intensities
r$\rightarrow$100) and the photospheric absorptions MgII, SiII exhibit
bona fide day-to-day velocity variations amounting to 7\,km/s.

Shallow FeI, FeII, and TiII absorptions and MgI, MgII lines in the
4280--4700\,\AA{} wavelength interval yield a spectral type of A2.7$\pm$0.3~Ib.
The equivalent widths and profiles of H$\delta$ and H$\gamma$, and the
equivalent width of the OI\,7774\,\AA{} triplet yield an absolute magmnitude
of Mv\,=\,$-5.5\pm  0.3^{\rm m}$.

\newpage

\begin{figure}[t]	      		      
\includegraphics[angle=-90,width=1.0\textwidth,bb=40 140 570 780,clip]{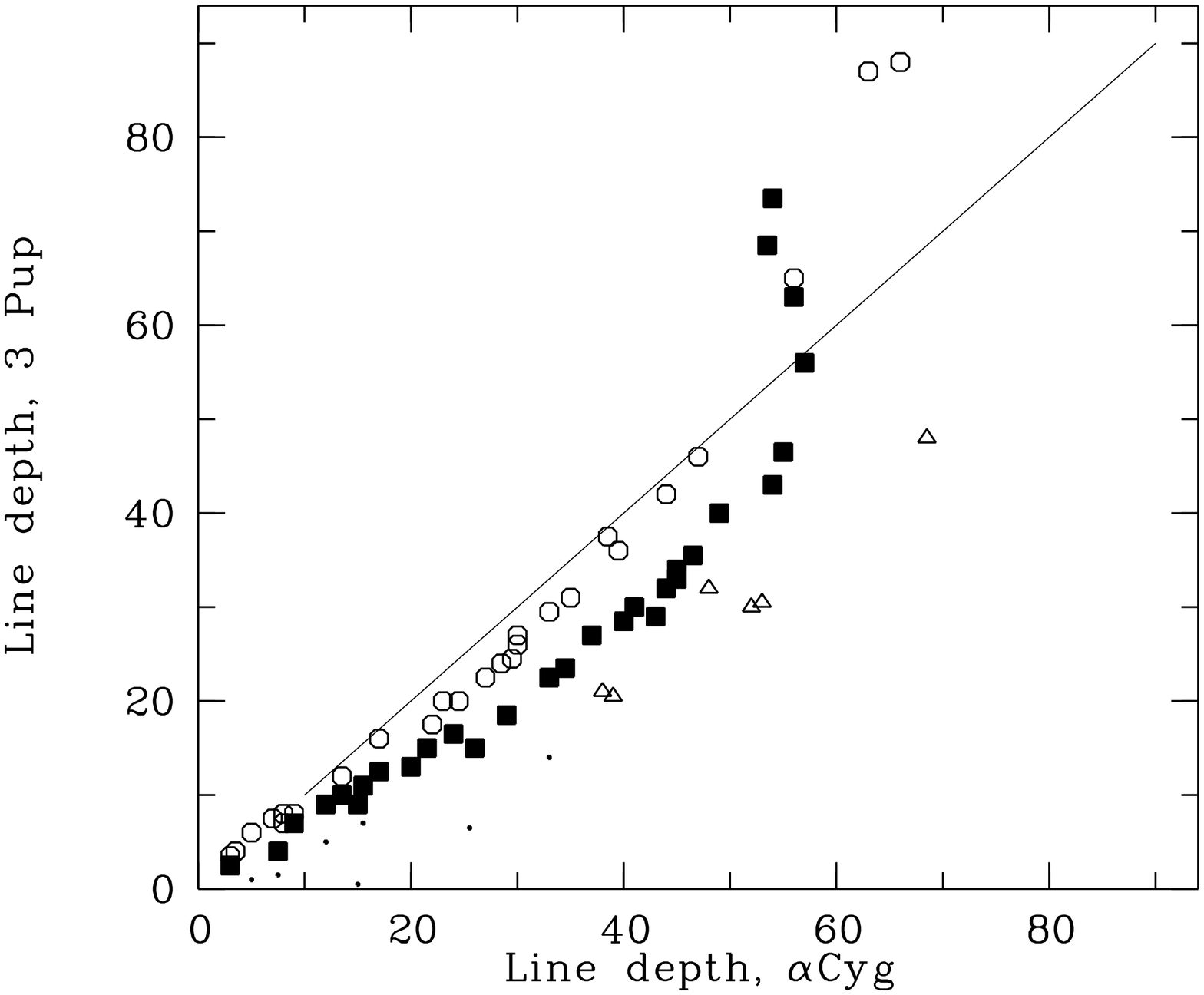}
\caption{Comparison of the central line depths R in the spectra of 3\,Pup
  (the vertical axis) and $\alpha$Cyg (the horizontal axis). The open
  circles correspond to Ti II lines in the 3680--5190\,\AA{} wavelength
  interval; the filled squares, to the FeII lines in the 
  3780--5170\,\AA{} wavelength interval; the dots, to the FeII lines in the
  5990--6520\,\AA{} wavelength interval, and the triangles, to the MgII(4)
and SiII(1,2,3) lines.}
\end{figure}

\newpage

\begin{figure}[t]	      		      
\includegraphics[angle=-90,width=1.0\textwidth,bb=40 40 560 780,clip]{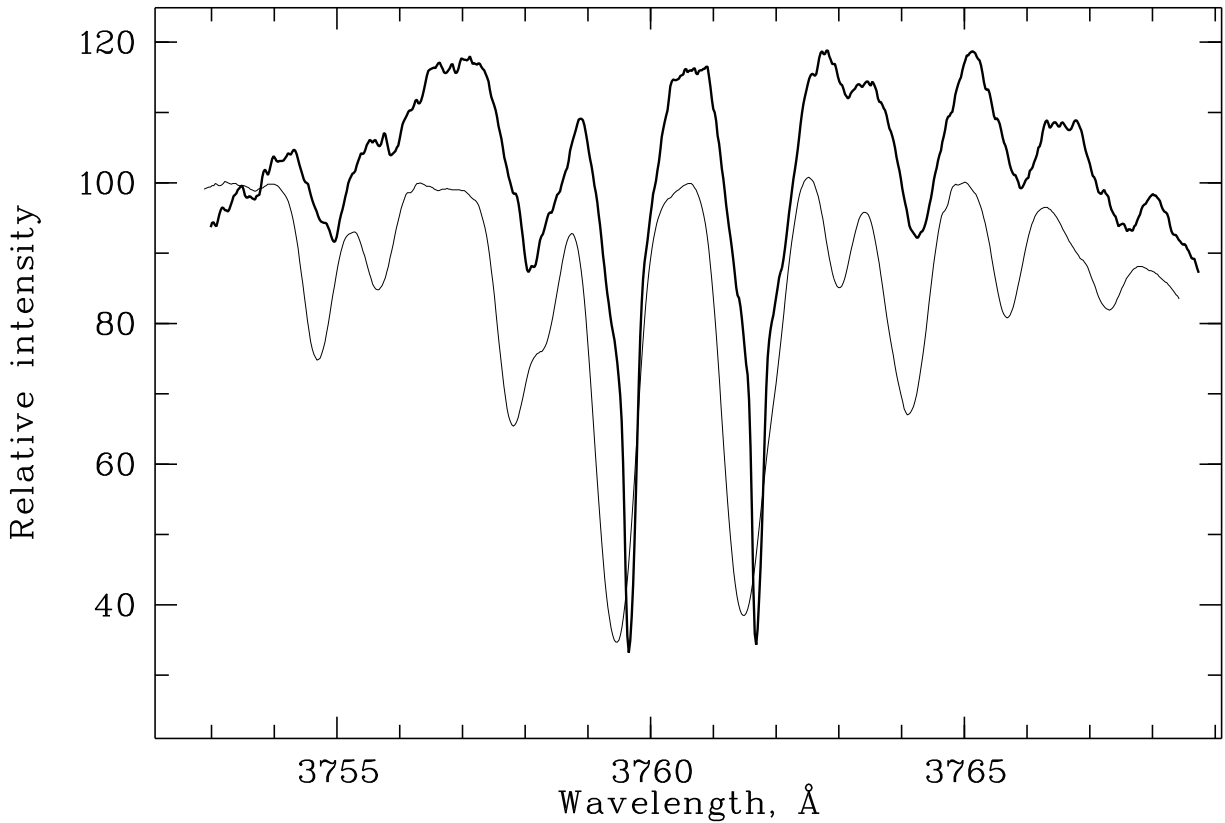}
\caption{Spectra of $\alpha$Cyg (the thin line at the bottom) and 3\,Pup
  (shifted upward by 20 scale points). The strongest lines are TiII\,(13)
  3759 and 3761\,\AA{}. The spectrum of 3\,Pup at the borders of the figure is
  depressed by the wings of the H12~3750 and H11~3771\,\AA{} lines. Here and
  in the subsequent figures the relative intensity r\,=\,100 corresponds to
  the continuum level.}
\end{figure}

\newpage

\begin{figure}[t]
\includegraphics[angle=0,width=0.5\textwidth,bb=50 60 550 770,clip]{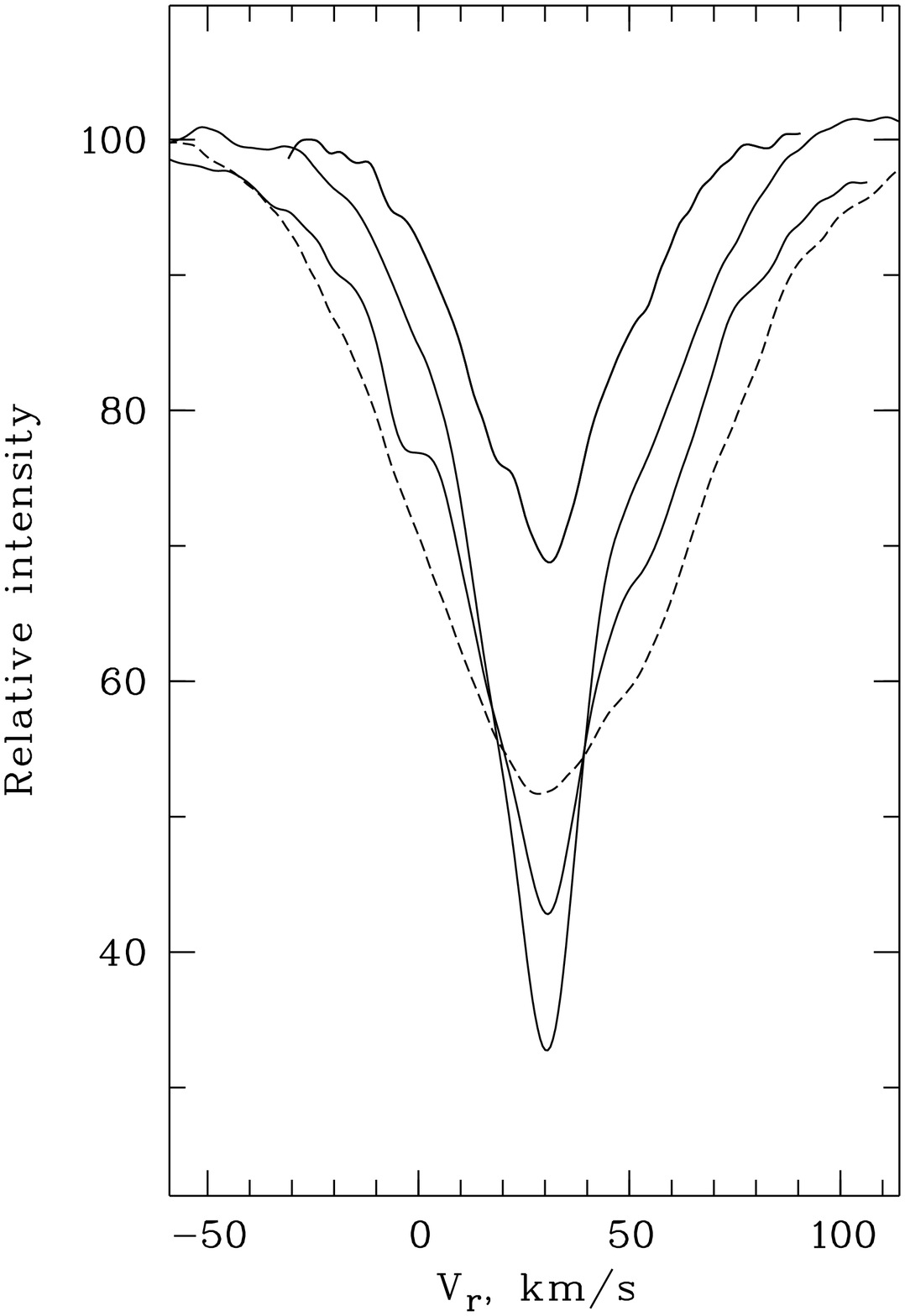}
\includegraphics[angle=0,width=0.5\textwidth,bb=50 60 550 770,clip]{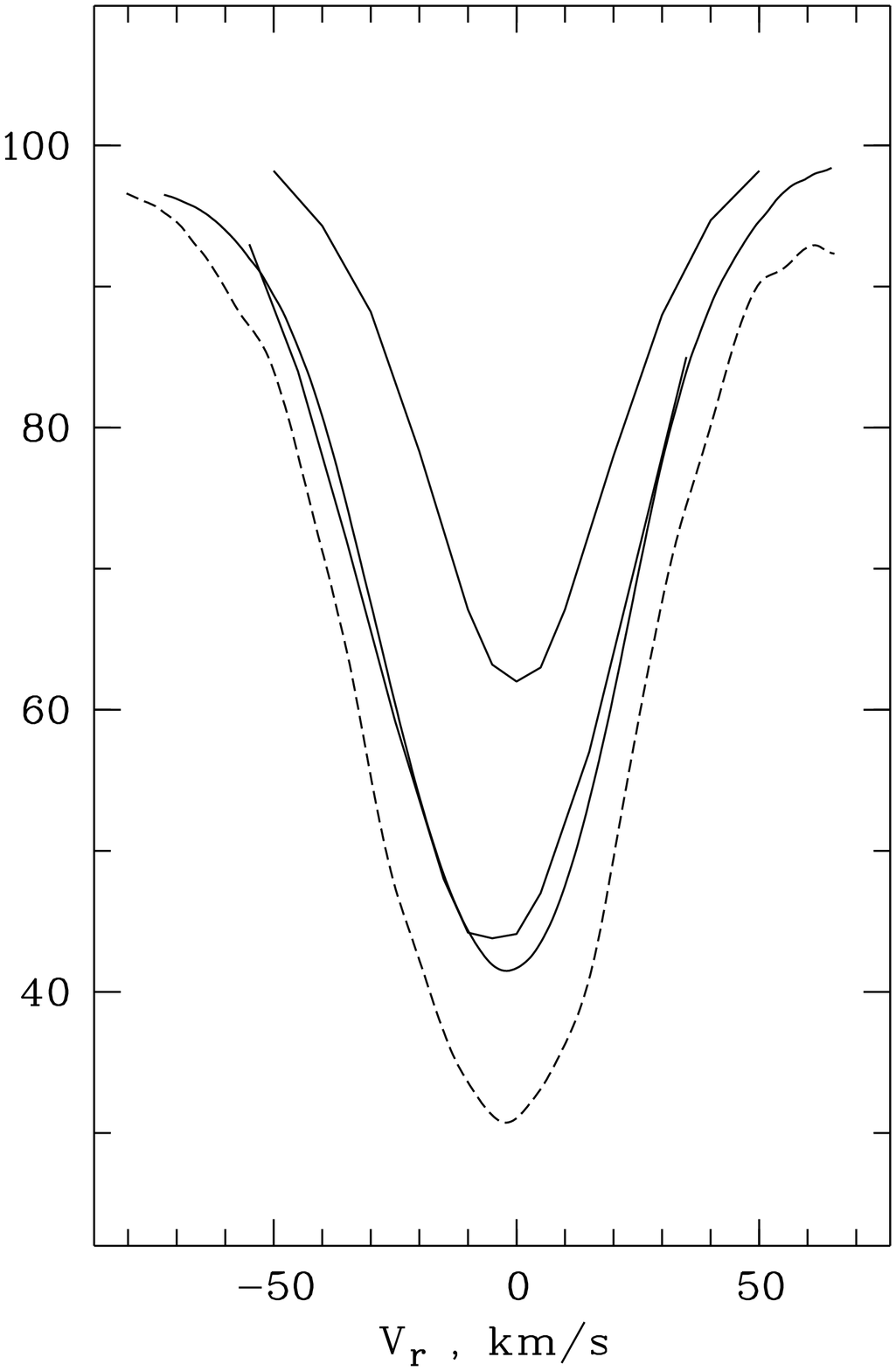}
\caption{Profiles of the FeII and MgII\,(4) lines in the spectra of
   3\,Pup (December, 26, 2006, on the left) and $\alpha$\,Cyg (on the right).
   The solid lines (from top to bottom) show the profiles of the following
   lines: FeII(49)~5235, FeII(27)~4233, and FeII(42)~4924\,\AA{}. The dashed
   lines show the profiles of MgII4481\,\AA{}.}
\end{figure}

\newpage

\begin{figure}[t]	      		      
\includegraphics[angle=0,width=0.6\textwidth,bb=80 80 550 770,clip]{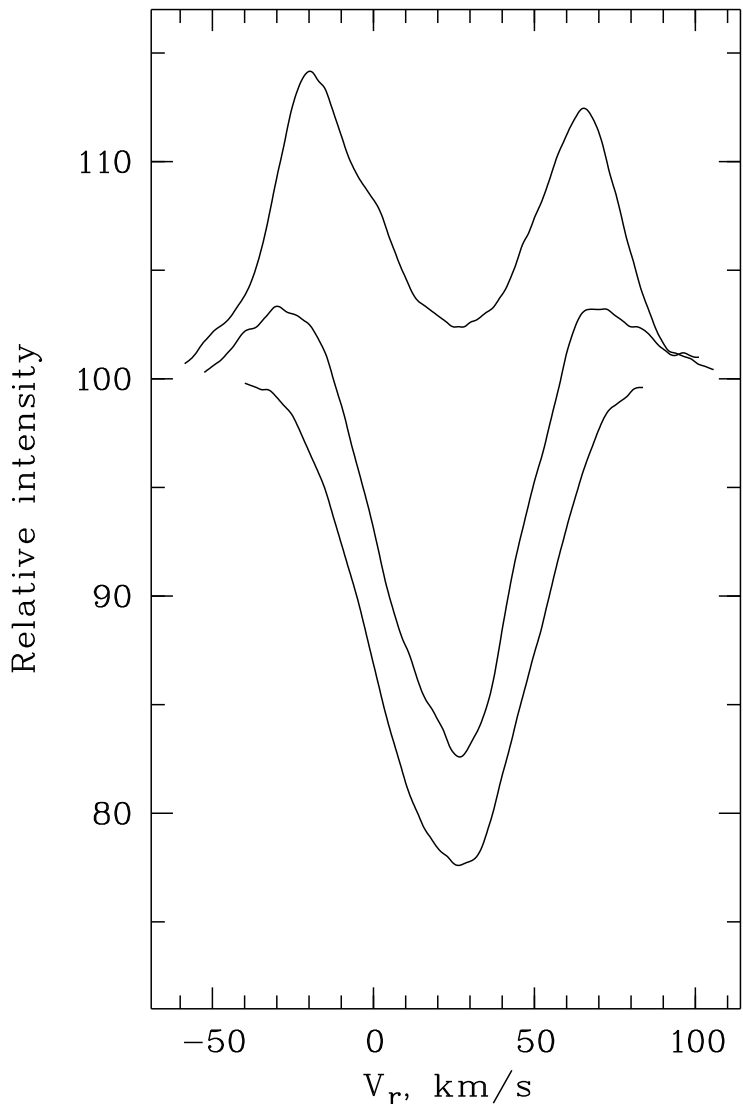}
\caption{Evolution of the FeII line profile with wavelength in the
       spectrum of 3\,Pup taken on December 25, 2004. From top to bottom:
       FeII(28)~4297, FeII(48)~5363, FeII(40)~6433\,\AA{}.}
\end{figure}

\newpage

\begin{figure}[t]	      		      
\includegraphics[angle=0,width=0.6\textwidth,bb=80 80 550 770,clip]{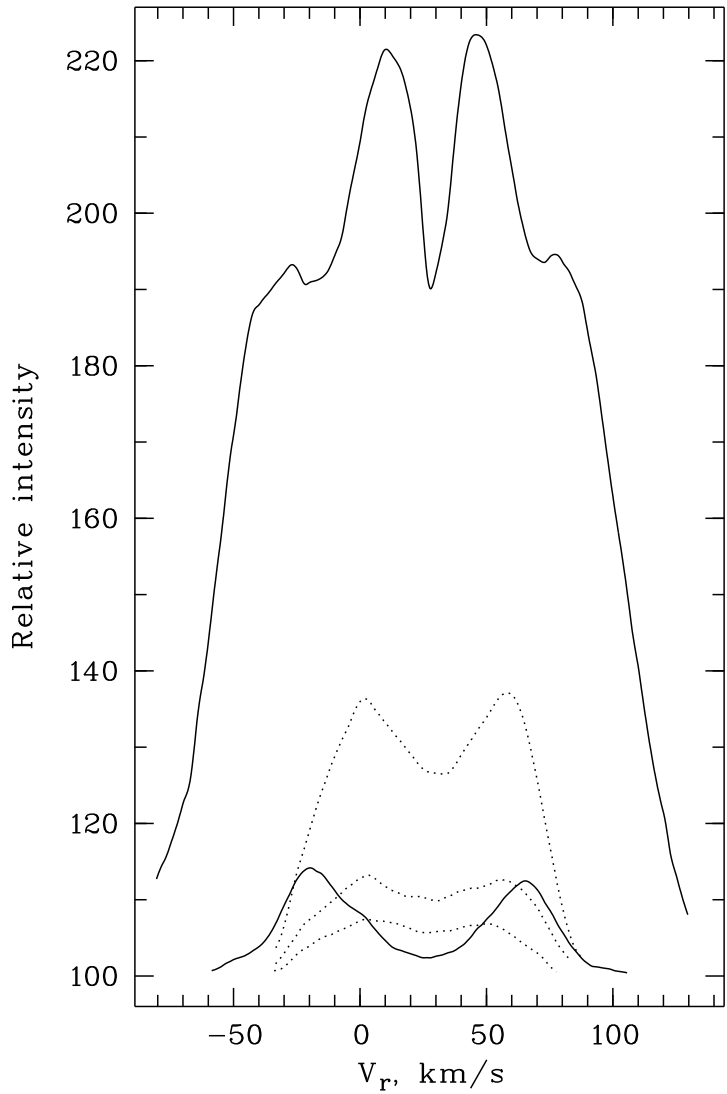}
\caption{Non-hydrogen emission features in the spectrum of 3\,Pup taken on
      December 25, 2004. The solid lines show the profiles of permitted lines
     (from top to bottom) CaII(2)~8542 and FeII(40)~6433\,\AA{}. The dashed
     lines (from top to bottom) show the profiles of the forbidden emissions
     [CaII]\,1F~7324, [OI]\,1F~6300, and [FeII]\,14F~7155\,\AA{}.}
\end{figure}

\newpage

\begin{figure}[t]	      		      
\includegraphics[angle=0,width=0.5\textwidth,bb=50 70 550 770,clip]{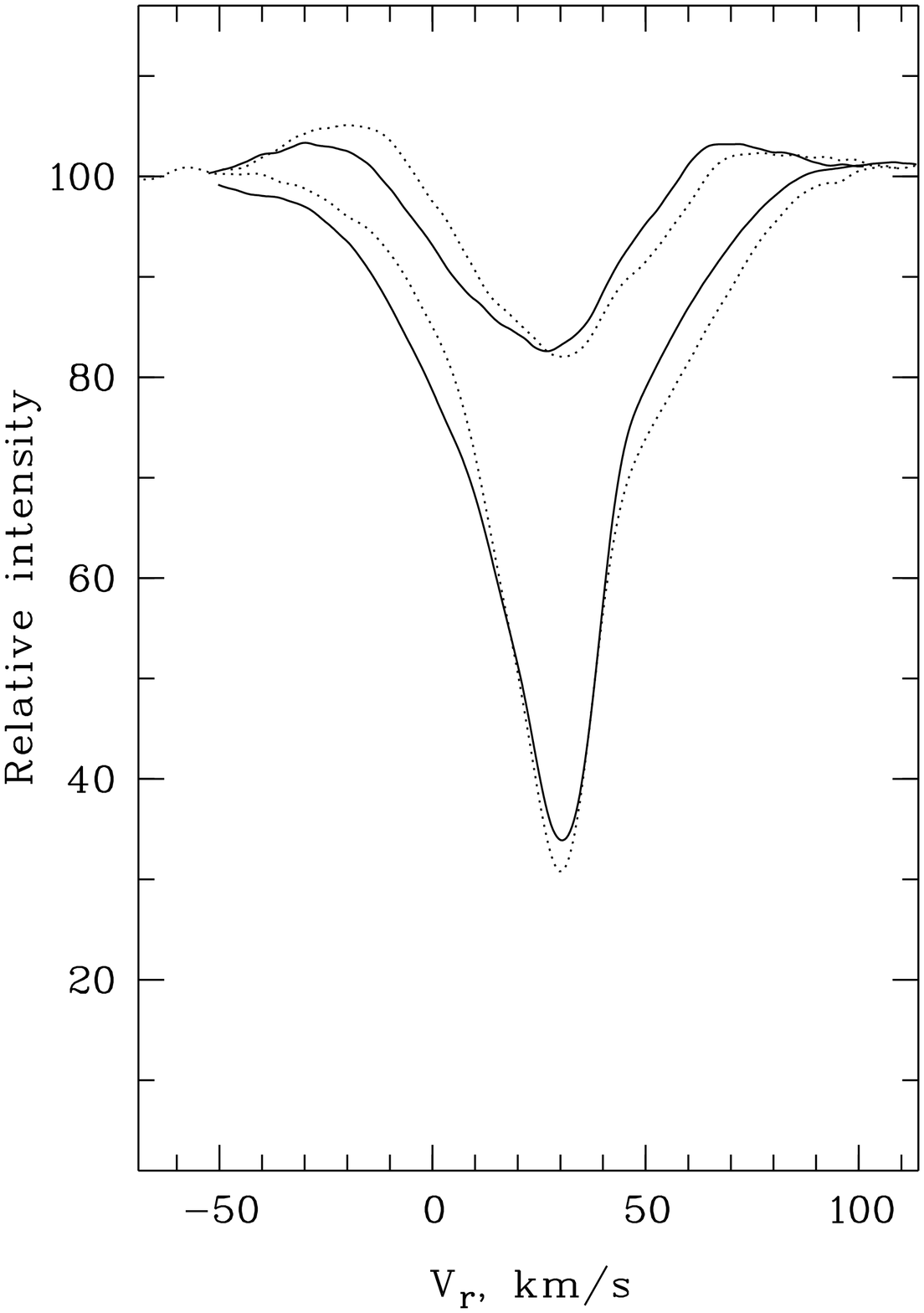}
\includegraphics[angle=0,width=0.5\textwidth,bb=50 70 550 770,clip]{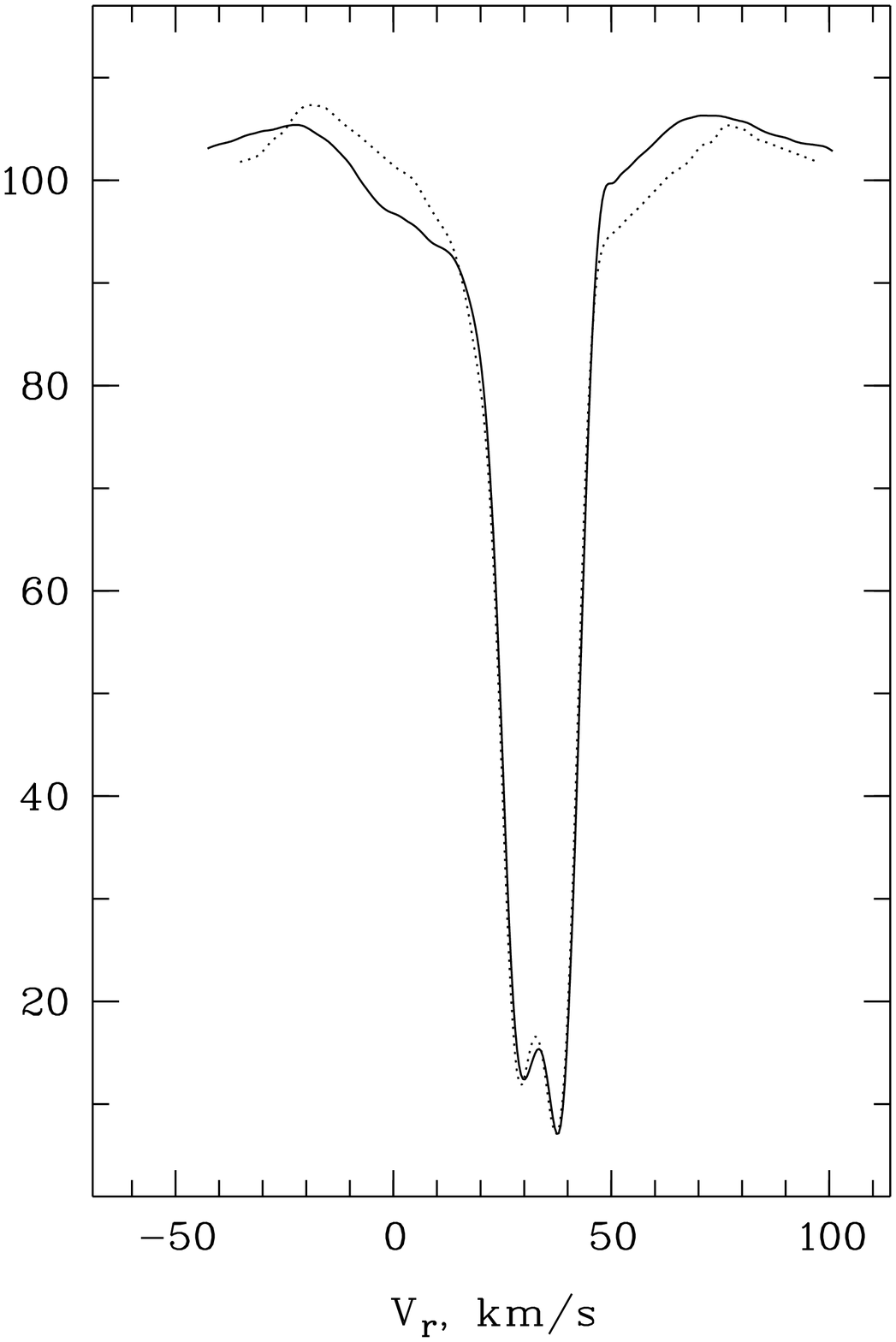}
\caption{Line profile variations in the spectrum of 3\,Pup. The solid and
      dotted lines correspond to the spectra taken on December 25, 2004
      and December 26, 2006, respectively. Left: FeII(48)~5363\,\AA (top) and
      averaged FeII(42)~4924, 5018\,\AA{}. Right: NaI(1)~5896\,\AA{}.}
\end{figure}

\newpage

\hoffset=-2cm
\begin{figure}[t]	      		      
\includegraphics[angle=-90,width=1.1\textwidth,bb=30 30 560 770,clip]{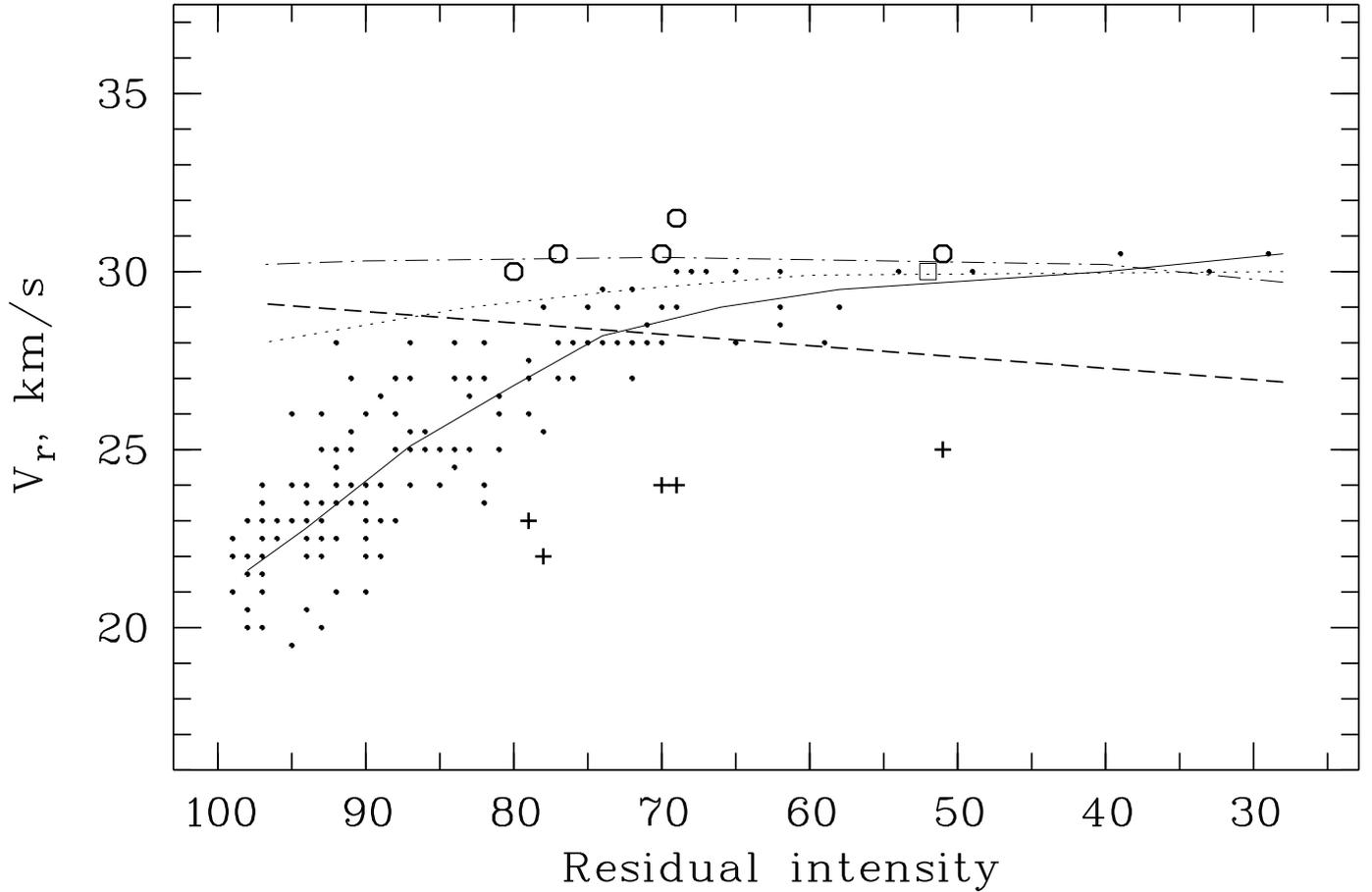}
\caption{Dependence of the radial velocity measured from the absorption
       core on its central intensity. The dots show individual absorptions of
       iron-group ions in the spectrum taken on December 25, 2004 and the solid
       line shows the Vr(r) curve averaged over these absorptions. The mean
       Vr(r) dependencies for the same group of absorptions are shown by: the
       dotted line (14.03.03), the dashed line (4.11.2008), and the dash-and-dot
       line (26.12.06 and 07.02.02). The MgII and SiII lines are shown by the
       crosses (the spectrum of December 25, 2004), open circles (December 26,
       2006), and the square (4.11.2008).}
\end{figure}

\newpage

\hoffset=2cm
\begin{figure}[t]	      		      
\includegraphics[angle=0,width=0.7\textwidth,bb=70 40 555 770,clip]{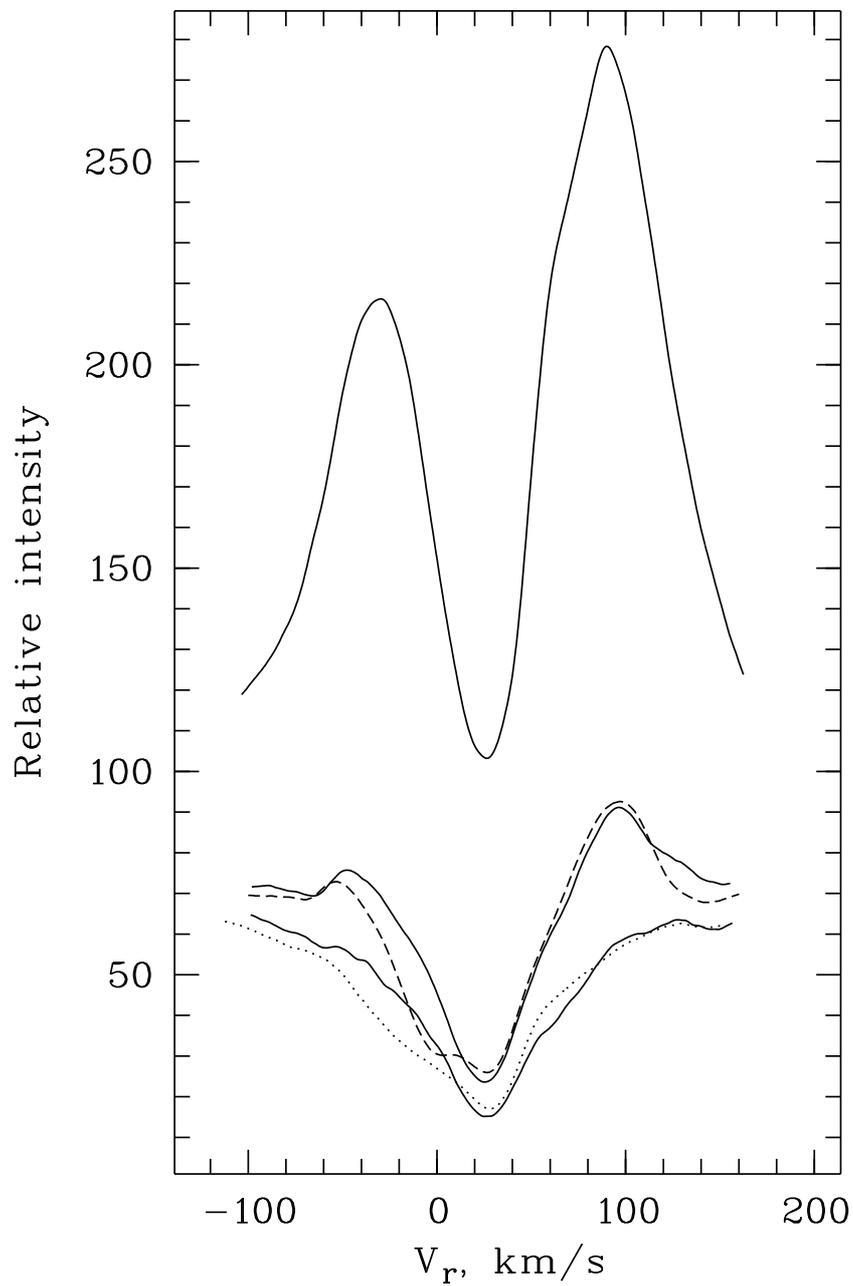}
\caption{Central parts of Balmer lines in the spectrum of 3\,Pup taken on
        December 26, 2006. From top to bottom: H$\alpha$, H$\beta$, and H$\gamma$.
	For comparison, the dashed and dotted lines show the spectrum of H$\beta$
	taken on November 4, 2008 and the spectrum of H$\gamma$ taken on April 125.12.04.}
\end{figure}

\newpage
\hoffset=-1cm
\begin{figure}[t]	      		      
\includegraphics[angle=0,width=0.6\textwidth,bb=30 30 560 770,clip]{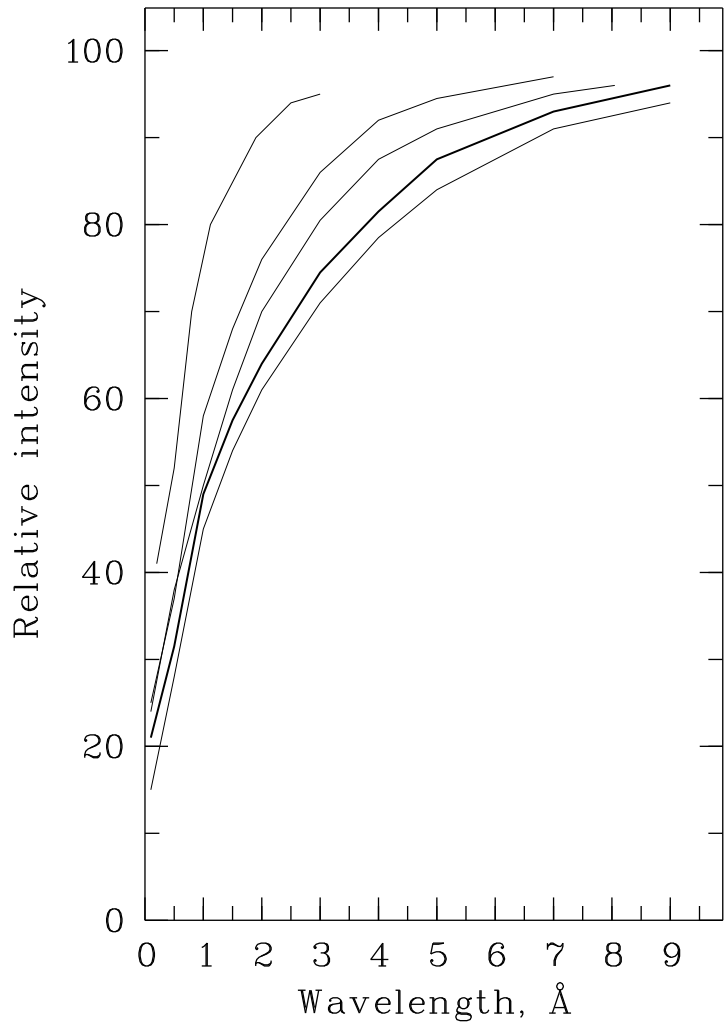}
\caption{H$\gamma$ half profiles in the spectra of 3\,Pup (the thick line) and
        of the comparison stars (the thin lines, from top to bottom: 6\,Cas~(A3\,Ia-0),
	$\alpha$\,Cyg~(A2\,Ia), HD\,13476~(A3\,Iab), HD\,210221~(A3\,Ib).}
\end{figure}

\clearpage
\newpage

\begin{table}[t]
%\setcaptionmargin{0mm}
\bigskip
\caption{Log of the spectra of 3\,Pup used in this paper. The last column
         gives some details concerning the telluric-line correction of Vr.}
\begin{tabular}{r|c|l|c|l}
\hline
 Date    &$\Delta\lambda$& Spectrograph   & $\lambda/\Delta\lambda$  &  Remarks \\
\hline
12.11.97 &   4690--8600 &  PFES  & 15000 &  for $\lambda >$5800\,\AA{} \\
 3.12.98 &   3940--5675 &  PFES  & 15000 &   нет \\ 
14.03.03 &   5160--6650 &  NES   & 60000 &  with errors of about  $\approx$0.2\,km/s \\
25.12.04 &   3700--8870 &  CFHT  &  70000&  with errors of about  $\approx$0.2\,km/s  \\
26.12.06 &   3650--8600 &  McDon &  60000&  with errors of about  $\approx$0.2\,km/s \\   
 7.02.07 &   4550--6010 &  NES   &  60000&  with errors of about  $\approx$0.2\,km/s \\  
 4.11.08 &   4460--5930 &  NES   &  60000&  with errors of about  $\approx$0.2\,km/s \\
\hline
\end{tabular}
\label{Data}
\end{table}

\clearpage
\newpage

\begin{table}[t]
\vspace{-0.5cm}
\caption{Heliocentric velocities inferred from individual lines and the
       mean velocities averaged over groups of lines in the spectra of 3\,Pup.
       For the NaI and FeII(42) lines observed on December 25, 2004 we list
       the velocities of the decoupled core components under the corresponding
       the mean velocities Vr.}
\begin{tabular}{ l | c|  c|  c|  c|  c|  c}
\hline
  &\multicolumn{6}{c}{\small V$_{\rm r}$, km/s} \\
\cline{2-7}
Dates  & 12.11.97 &  14.03.03 & 25.12.04 & 26.12.06 & 07.02.07 & 04.11.08 \\
\hline
\multicolumn{7}{l}{\underline{\small Interstellar lines:}} \\
NaI(1)          &    33.2  &     33.0  &   33.6   &   33.3   &    33.5 &  33.0  \\
                &     --   &     29.6  &   29.4   &   29.7   &    30.6:&  30.7: \\
                &     --   &     36.3  &   37.4   &   37.4   &    36.5 &  35.8  \\
KI(1)           &    28:   &      --   &   28     &   28.7:  &   --    &  --    \\
                &     --   &      --   &   37     &   --     &   --    &  --    \\
CaII(1)         &    --    &     --    &   27.0   &    28.1  &   --    &  --    \\ 
                &          &           &   38:    &  --      &   --    &  --    \\
DIB             &    30.5  &      31:  &   31.0   &   29.5   &    31   &   29   \\ 
\multicolumn{7}{l}{\underline{\small Stellar emissions:}} \\  
$[\rm OI]$      &    31:   &      27   &   28     &   29     &   --    &  --    \\
$[\rm FeII], [\rm CaII]$ &   25: &  -- &   28     &   28     &    30:  &   26:  \\
FeII et al      &    28:   &      26   &   24     &   27.5   &    28   &   27   \\
\multicolumn{7}{l}{\underline{\small Stellar absorptions:}} \\ 
FeII(42)        &    30    &      30:  &   30.5   &   29.8   &    29.4 &   27.3 \\
                &          &           &   28.8	  &	     &	       &        \\
FeII, TiII et al&    29.5: &      29.5:&   28.5   &   30.5   &    30.0 &   28.0 \\ 
MgII, SiII      &    27:   &      32   &   24     &   30.2   &    32:  &   30   \\
The weakest lines(r $\rightarrow 100$)& 28& 28 &   23     &   30.5   &    29.8 &   29.2 \\  
H$\delta$       &   --     &    --     &   28     &   29     &    --   &   --   \\      
H$\gamma$       &   --     &     --    &   28     &   26     &    --   &   --   \\
H$\beta$        &    28    &     --    &   28     &   25     &     27  &    26  \\ 
H$\alpha$       &    24    &      26   &   22     &   26     &    --   &   --   \\ 
\hline
\end{tabular}
\label{Velocity}
\end{table}

\newpage
\clearpage

\tablecaption{Residual intensities r and heliocentric radial velocities Vr for
        individual lines in the high-resolution spectra of 3\,Pup, taken on
        different dates.}
\tablehead{\hline 
        &  & \multicolumn{2}{c|}{14.03.03}&\multicolumn{2}{c|}{25.12.04} 
       &\multicolumn{2}{c|}{26.12.06} &\multicolumn{2}{c|}{07.02.07}
       &\multicolumn{2}{c|}{04.11.08} \\
       \cline{3-12} \rule{0pt}{5pt} Element& $\lambda$,\AA{}  & $r$ & $V_{\odot}$  & $r$ & $V_{\odot}$ & $r$ & $V_{\odot}$ & $r$ & $V_{\odot}$ & $r$ & $V_{\odot}$ \\
       \hline } 
\tabletail{ \hline}        
\begin{supertabular}{|lc|  cc|  cc|  cc|  cc|  cc|}
H20         & 3682.81&          &     &              &    &    0.52&  31&        &    &        &     \\ 
TiII(14)    & 3685.19&          &     &              &    &    0.32&  31&        &    &        &     \\                                                       
H19         & 3686.83&          &     &              &    &    0.50&  35&	 &    &	       &     \\ 
H18         & 3691.56&          &     &              &    &    0.40&  34&	 &    &	       &     \\ 
H17         & 3697.15&          &     &              &    &    0.38&  31&	 &    &	       &     \\ 
H16         & 3703.85&          &     &              &    &    0.37&  34&	 &    &	       &     \\ 
CaII(3)     & 3706.02&          &     &              &    &    0.57&  26&	 &    &	       &     \\ 
H15         & 3711.97&          &     &              &    &    0.34&	&	 &    &	       &     \\ 
H14         & 3721.94&          &     &              &    &    0.35&  35&	 &    &	       &     \\ 
VII(21)     & 3727.35&          &     &              &    &    0.72&  32&	 &    &	       &     \\ 
H13         & 3734.37&          &     &          0.32&  30&    0.28&  35&	 &    &	       &     \\ 
CaII(3)     & 3736.89&          &     &              &    &    0.51&  33&     	 &    &	       &     \\ 
CrII(20)    & 3738.38&          &     &              &    &    0.73&  27&	 &    &	       &     \\ 
TiII(72)    & 3741.64&          &     &          0.62&  29&    0.59&  31&	 &    &	       &     \\ 
H12         & 3750.15&          &     &          0.29&    &        &  28&	 &    &	       &     \\ 
CrII(20)    & 3754.56&          &     &              &  31&        &  33&	 &    &	       &     \\ 
TiII(13)    & 3759.29&          &     &          0.13&  29&    0.09&  30&	 &    &	       &     \\ 
TiII(13)    & 3761.32&          &     &          0.13&  28&    0.13&  31&	 &    &	       &     \\ 
FeI(608)    & 3765.54&          &     &          0.79&  30&	   &	&	 &    &	       &     \\ 
FeI(21)     & 3767.20&          &     &              &    &        &  29&	 &    &	       &     \\ 
H11         & 3770.63&          &     &          0.24&  25&    0.24&  34&	 &    &	       &     \\ 
TiII(72)    & 3776.05&          &     &              &  22&        &  30&	 &    &	       &     \\ 
VII(21)     & 3778.36&          &     &              &  23&	   &	&	 &    &	       &     \\ 
FeII(14)    & 3783.34&          &     &          0.72&  29&    0.73&  34&	 &    &	       &     \\ 
TiII(12)    & 3786.33&          &     &          0.85&  25&	   &	&	 &    &	       &     \\ 
H10         & 3797.90&          &     &          0.22&  24&    0.23&  31&	 &    &	       &     \\ 
FeI(45)     & 3815.84&          &     &              &    &    0.78&  28&	 &    &	       &     \\ 
FeI(20)     & 3820.42&          &     &          0.78&  24&    0.77&  32&	 &    &	       &     \\ 
FeI(20)     & 3825.88&          &     &          0.79&  26&    0.81&  33&	 &    &	       &     \\ 
MgI(3)      & 3829.36&          &     &          0.75&  27&    	   &	&	 &    &	       &     \\ 
H9          & 3835.38&          &     &          0.21&  26&        &  33&	 &    &	       &     \\ 
MgI(3)      & 3838.29&          &     &              &    &        &  27&	 &    &	       &     \\ 
FeI(124)    & 3845.17&          &     &          0.84&  27&	   &	&	 &    &	       &     \\ 
SiII(1)     & 3853.66&          &     &          0.79&  23&    0.80&  35&	 &    &	       &     \\ 
SiII(1)     & 3856.02&          &     &          0.59&    &    0.58&  35&	 &    &	       &     \\ 
FeI(4)      & 3859.91&          &     &          0.81&  27&    0.82&  31&	 &    &	       &     \\ 
SiII(1)     & 3862.60&          &     &          0.70&  24&    0.69&  33&	 &    &	       &     \\ 
CrII(167)   & 3865.59&          &     &          0.82&  24&    0.82&  31&	 &    &	       &     \\ 
FeI(4)      & 3878.58&          &     &              &    &    0.74&  30&	 &    &	       &     \\ 
TiII(34)    & 3882.28&          &     &              &    &    0.84&  30&	 &    &	       &     \\ 
H8          & 3889.05&          &     &          0.21&  29&    0.18&  31&	 &    &	       &     \\    
TiII(34)    & 3900.54&          &     &          0.55&  28&    0.53&  30&	 &    &	       &     \\ 
TiII(34)    & 3913.46&          &     &          0.59&  28&    0.58&  31&	 &    &	       &     \\ 
VII(10)     & 3916.40&          &     &          0.92&  22&    0.92&  33&	 &    &	       &     \\ 
FeI(4)      & 3922.91&          &     &          0.91&  24&	   &	&	 &    &	       &     \\ 
FeI(4)      & 3927.92&          &     &          0.93&  22&    0.91&  32&	 &    &	       &     \\ 
FeI(4)      & 3930.30&          &     &          0.85&  25&    0.83&  27&	 &    &	       &     \\ 
CaII(1)     & 3933.66&          &     &          0.02&  27&    0.01&  28&	 &    &	       &     \\ 
FeII(173)   & 3935.94&          &     &          0.87&  24&	   &	&	 &    &	       &     \\ 
AlI(1)      & 3944.01&          &     &          0.91&  21&	   &	&	 &    &	       &     \\ 
FeII(3)     & 3945.21&          &     &          0.89&  22&    0.88&  27&   	 &    &	       &     \\ 
VII(10)     & 3951.96&          &     &          0.91&  27&    0.90&  34&	 &    &	       &     \\ 
FeI(278)    & 3956.68&          &     &              &    &    0.94&  30&	 &    &	       &     \\ 
Al(1)       & 3961.52&          &     &              &  19&	   &	&	 &    &	       &     \\ 
CaII(1)     & 3968.47&          &     &              &  27&        &  28&	 &    &	       &     \\ 
H$\epsilon$ & 3970.07&          &     &          0.20&  31&    0.18&  32&	 &    &	       &     \\ 
CrII(183)   & 3979.52&          &     &          0.89&  24&    0.88&  30&	 &    &	       &     \\ 
TiII(11)    & 3981.99&          &     &              &    &    0.89&  30&    	 &    &	       &     \\ 
TiII(11)    & 3987.61&          &     &          0.95&  19&	   &	&	 &    &	       &     \\ 
VII(9)      & 3997.11&          &     &          0.93&  27&    0.91&  33&	 &    &	       &     \\ 
VII(32)     & 4023.38&          &     &          0.91&  23&    0.91&  33&	 &    &	       &     \\  
HeI(18)     & 4026.22&          &     &              &    &    0.93&  30&	 &    &	       &     \\ 
TiII(87)    & 4028.34&          &     &          0.83&  27&    0.85&  30&	 &    &	       &     \\ 
FeII(126)   & 4032.95&          &     &          0.90&  23&    0.90&  30&	 &    &	       &     \\ 
VII(32)     & 4035.61&          &     &          0.92&  25&    0.91&  29&    	 &    &	       &     \\ 
FeI (43)    & 4045.81&          &     &          0.83&  24&    0.83&  32&	 &    &	       &     \\ 
FeII(172)   & 4048.83&          &     &              &    &    0.90&  32&	 &    &	       &     \\ 
CrII(19)    & 4051.97&          &     &          0.90&  21&    0.91&  31&	 &    &	       &     \\ 
TiII(87)    & 4053.83&          &     &              &    &    0.81&  31&	 &    &	       &     \\ 
FeI(43)     & 4063.59&          &     &          0.86&  24&        &	&	 &    &	       &     \\ 
NiII(11)    & 4067.03&          &     &              &    &    0.84&  32&	 &    &	       &     \\ 
FeI(43)     & 4071.74&          &     &              &    &    0.88&  30&	 &    &	       &     \\ 
SrII(1)     & 4077.72&          &     &          0.85&  21&    0.85&  26&	 &    &	       &     \\ 
H$\delta$   & 4101.74&          &     &          0.17&  28&    0.15&  29&   	 &    &	       &     \\ 
CrII(18)    & 4110.99&          &     &          0.90&  22&    0.92&  30&	 &    &	       &     \\ 
FeII(28)    & 4122.66&          &     &          0.84&  25&    0.86&  30&   	 &    &	       &     \\ 
SiII(3)     & 4130.88&          &     &          0.78&  22&    0.80&  30&	 &    &	       &     \\ 
TiII(21)    & 4161.52&          &     &              &    &    0.90&  30&   	 &    &	       &     \\ 
TiII(105)   & 4163.64&          &     &          0.81&  26&    0.79&  31&	 &    &	       &     \\ 
MgI(15)     & 4167.27&          &     &          0.97&  21&	   &	&	 &    &	       &     \\ 
TiII(105)   & 4171.90&          &     &              &    &    0.82&  32&	 &    &	       &     \\ 
FeII(27)    & 4173.46&          &     &          0.67&  30&    0.67&  32&   	 &    &	       &     \\ 
FeII(21)    & 4177.68&          &     &          0.82&  23&    0.84&  30&	 &    &	       &     \\ 
FeII(32)    & 4178.85&          &     &          0.65&  30&    0.64&  31&	 &    &	       &     \\ 
FeI(152)    & 4187.80&          &     &              &    &    0.93&  30&	 &    &	       &     \\ 
CrII(161)   & 4195.41&          &     &          0.96&  23&	   &	&	 &    &	       &     \\ 
SrII(1)     & 4215.52&          &     &          0.89&  27&    0.88&  33&    	 &    &	       &     \\ 
FeII(27)    & 4233.17&          &     &          0.48&  30&    0.43&  31&        &    &	       &     \\ 
CrII(31)    & 4242.37&          &     &          0.79&  27&    0.79&  32&        &    &	       &     \\ 
ScII(7)     & 4246.83&          &     &          0.76&  28&    0.79&  32&	 &    &	       &     \\ 
CrII(31)    & 4252.63&          &     &          0.90&  26&        &	&	 &    &	       &     \\ 
VII(18)     & 4254.41&          &     &          0.93&  23&    0.93&  28&	 &    &	       &     \\ 
FeII(28)    & 4258.15&          &     &          0.84&  27&    0.82&  32&	 &    &	       &     \\ 
FeI(152)    & 4260.47&          &     &          0.91&  24&    0.93&  32&	 &    &	       &     \\ 
CrII(31)    & 4261.92&          &     &          0.84&  25&    0.84&  29&	 &    &	       &     \\ 
CrII(31)    & 4269.29&          &     &          0.93&  20&    0.92&  28&   	 &    &	       &     \\ 
FeII(27)    & 4273.32&          &     &          0.84&  25&    0.83&  31&	 &    &	       &     \\ 
FeII(32)    & 4278.15&          &     &          0.91&  25&    0.91&  31&	 &    &	       &     \\ 
CrII(31)    & 4284.20&          &     &          0.88&  23&    0.87&  30&   	 &    &	       &     \\ 
TiII(41)    & 4290.21&          &     &          0.76&  27&    0.75&  29&   	 &    &	       &     \\ 
TiII(20)    & 4294.10&          &     &          0.73&  28&    0.73&  31&	 &    &	       &     \\ 
FeII(28)    & 4296.57&          &     &          0.77&  27&    0.78&  30&	 &    &	       &     \\ 
TiII(41)    & 4300.04&          &     &          0.65&  28&    0.63&  31&	 &    &	       &     \\ 
TiII(41)    & 4301.92&          &     &          0.82&  28&    0.83&  31&	 &    &	       &     \\ 
FeII(27)    & 4303.17&          &     &          0.69&  30&    0.67&  31&	 &    &	       &     \\ 
TiII(41)    & 4307.89&          &     &          0.78&  25&    0.77&  29&	 &    &	       &     \\ 
TiII(41)    & 4312.86&          &     &          0.79&  27&    0.81&  30&	 &    &	       &     \\ 
TiII(94)    & 4316.80&          &     &          0.93&  23&    0.92&  32&	 &    &	       &     \\ 
H$\gamma$   & 4340.47&          &     &          0.17&  28&    0.15&  26&	 &    &	       &     \\ 
FeII(27)    & 4351.77&          &     &          0.58&  29&    0.56&  29&	 &    &	       &     \\ 
FeII(213)   & 4354.36&          &     &          0.95&  26&    0.95&  32&	 &    &	       &     \\ 
FeII        & 4357.57&          &     &          0.94&  21&    0.93&  31&	 &    &	       &     \\ 
$[$FeII$]$7F& 4359.33&          &     &          1.02&	  &	   &	&	 &    &	       &     \\ 
FeII(28)    & 4369.40&          &     &          0.89&  23&    0.90&  29&	 &    &	       &     \\ 
FeII(27)    & 4385.38&          &     &          0.72&  27&    0.71&  28&	 &    &	       &     \\ 
TiII(104)   & 4386.85&          &     &          0.92&  22&	   &	&	 &    &	       &     \\ 
TiII(51)    & 4395.03&          &     &          0.62&  28&    0.63&  31&	 &    &	       &     \\ 
TiII(51)    & 4399.77&          &     &          0.82&  28&	   &	&	 &    &	       &     \\ 
FeI (41)    & 4404.75&          &     &          0.88&  22&    0.89&  32&	 &    &	       &     \\ 
FeII(32)    & 4413.59&          &     &          0.97&  23&    0.95&  32&	 &    &	       &     \\ 
FeII(27)    & 4416.82&          &     &          0.75&  29&    0.73&  30&	 &    &	       &     \\ 
TiII(93)    & 4421.94&          &     &          0.96&  27&    0.96&  31&	 &    &	       &     \\ 
MgII(9)     & 4433.99&          &     &          0.95&  21&    0.95&  30&	 &    &	       &     \\ 
TiII(19)    & 4443.80&          &     &          0.70&  28&    0.71&  31&	 &    &	       &     \\ 
TiII(19)    & 4450.48&          &     &          0.86&  25&    0.85&  31&	 &    &	       &     \\ 
TiII(40)    & 4464.45&          &     &          0.90&  23&    0.89&  32&        &    &    0.92&  30 \\ 
FeI(350)    & 4466.55&          &     &              &    &        &    &        &    &    0.97&  28 \\ 
TiII(31)    & 4468.49&          &     &          0.72&  28&    0.71&  31&        &    &    0.73&  28 \\ 
HeI(14)     & 4471.52&          &     &              &    &        &    &        &    &    0.90&     \\ 
FeII(37)    & 4472.92&          &     &          0.90&  23&        &    &        &    &    0.90&  28 \\ 
FeI(350)    & 4476.04&          &     &          0.98&  23&        &    &        &    &    0.98&  27 \\ 
MgII(4)     & 4481.22&          &     &          0.51&  25&    0.52&  30&        &    &    0.52&  30 \\ 
FeII(37)    & 4489.17&          &     &              &    &    0.79&  29&        &    &    0.81&  29 \\ 
FeII(37)    & 4491.40&          &     &          0.77&  28&    0.76&  30&        &    &    0.76&  28 \\ 
TiII(18)    & 4493.52&          &     &              &    &        &    &        &    &    0.96&  31 \\ 
TiII(31)    & 4501.27&          &     &          0.74&  28&    0.74&  31&        &    &    0.76&  29 \\ 
FeII(38)    & 4508.28&          &     &          0.68&  30&    0.66&  32&        &    &    0.68&  28 \\ 
FeII(37)    & 4515.34&          &     &          0.71&  28&    0.69&  30&        &    &    0.71&  27 \\ 
TiII(18)    & 4518.36&          &     &          0.97&  23&    0.95&  31&	 &    &	       &     \\ 
FeII(37)    & 4520.22&          &     &          0.73&  29&    0.71&  30&        &    &    0.72&  27 \\ 
FeII(38)    & 4522.63&          &     &          0.62&  30&    0.59&  30&        &    &    0.60&  27 \\ 
TiII(82)    & 4529.49&          &     &              &    &    0.93&  31&        &    &        &     \\ 
CrII(39)    & 4539.62&          &     &          0.97&  24&        &    &        &    &        &     \\ 
FeII(38)    & 4541.52&          &     &          0.81&  26&    0.81&  29&        &    &    0.82&  28 \\ 
FeII(37)    & 4555.89&          &     &              &    &        &    &        &    &    0.67&  27 \\ 
CrII(44)    & 4558.65&          &     &          0.70&  29&        &    &    0.64&  31&    0.69&  28 \\ 
TiII(50)    & 4563.76&          &     &          0.76&  28&        &    &    0.76&  32&    0.75&  28 \\ 
CrII(39)    & 4565.77&          &     &          0.94&  22&        &    &    0.92&  30&    0.93&  29 \\ 
TiII(60)    & 4568.32&          &     &              &    &        &    &        &    &    0.98&  28 \\ 
TiII(82)    & 4571.97&          &     &          0.69&  29&        &    &    0.70&  30&    0.69&  29 \\ 
FeII(38)    & 4576.33&          &     &          0.82&  27&        &    &    0.80&  29&    0.81&  28 \\ 
FeII(26)    & 4580.06&          &     &          0.92&  21&        &    &    0.90&  26&    0.91&  27 \\ 
FeII(38)    & 4583.83&          &     &          0.54&  30&        &    &    0.49&  30&    0.50&  27 \\ 
CrII(44)    & 4588.20&          &     &          0.74&  28&        &    &    0.72&  31&    0.73&  28 \\ 
TiII(50)    & 4589.94&          &     &              &    &        &    &    0.88&  31&    0.88&  31 \\ 
CrII(44)    & 4592.05&          &     &          0.86&  25&        &    &    0.86&  32&    0.87&  29 \\ 
FeII(219)   & 4598.53&          &     &              &    &        &    &    0.97&  25&    0.98&  30 \\ 
VII(150)    & 4600.18&          &     &              &    &        &    &    0.97&  30&    0.98&  28 \\ 
CrII(44)    & 4616.62&          &     &          0.88&  25&        &    &    0.88&  30&    0.88&  30 \\ 
CrII(44)    & 4618.82&          &     &          0.79&  26&        &    &    0.79&  29&    0.78&  28 \\ 
FeII(38)    & 4620.51&          &     &          0.87&  25&        &    &    0.87&  31&    0.86&  29 \\  
SiII        & 4621.72&          &     &              &    &        &    &        &    &    0.98&  28 \\ 
FeII(186)   & 4625.91&          &     &          0.97&  23&        &    &    0.97&  29&    0.97&  30 \\ 
FeII(37)    & 4629.33&          &     &          0.70&  28&        &    &    0.69&  30&    0.69&  27 \\ 
CrII(44)    & 4634.07&          &     &              &    &        &    &    0.83&  31&    0.82&  30 \\ 
FeII(186)   & 4635.31&          &     &              &    &        &    &    0.89&  32&    0.90&  29 \\ 
FeI(822)    & 4638.02&          &     &              &    &        &    &    0.97&  29&    0.97&  29 \\ 
FeII        & 4640.84&          &     &              &    &        &    &        &    &    0.99&  29 \\ 
FeII(25)    & 4648.93&          &     &          0.98&  20&        &    &    0.97&  26&    0.98&  27 \\ 
FeII(37)    & 4666.75&          &     &          0.87&  25&        &    &    0.88&  30&    0.87&  29 \\ 
SiII        & 4673.27&          &     &              &    &        &    &        &    &    0.99&  29 \\ 
MgI (11)    & 4702.99&          &     &          0.96&  21&    0.96&  31&    0.96&  31&    0.96&  31 \\ 
TiII(49)    & 4708.67&          &     &          0.97&  20&    0.96&  29&    0.96&  27&    0.97&  30 \\ 
HeI(12)     & 4713.19&          &     &          0.98&  25&    0.97&  32&    0.98&  29&    0.98&  29 \\ 
$[$FeII$]$4F& 4728.07&          &     &          1.01&  24&    1.01&  24&    1.01&  22&    1.01&  23 \\ 
FeII(43)    & 4731.47&          &     &          0.85&  24&    0.85&  29&    0.85&  28&    0.86&  27 \\ 
MnII(5)     & 4755.72&          &     &          0.97&  24&        &    &    0.97&  30&    0.97&  32 \\ 
FeII(37)    & 4766.75&          &     &              &    &    0.87&  28&    	 &    &	       &     \\ 
TiII(92)    & 4779.98&          &     &          0.91&  25&    0.92&  31&    0.92&  30&    0.91&  31 \\ 
TiII(17)    & 4798.53&          &     &              &    &        &    &    0.98&  29&    0.98&  28 \\ 
TiII(92)    & 4805.09&          &     &          0.88&  27&        &    &    0.89&  30&    0.88&  29 \\ 
MnII(5)     & 4806.86&          &     &              &    &        &    &        &    &    0.99&  27 \\ 
CrII(30)    & 4812.35&          &     &          0.94&  23&    0.93&  30&    0.94&  31&    0.94&  30 \\ 
$[$FeII$]$20F& 4814.53&         &     &          1.02&  23&    1.01&  28&    1.02&  28&    1.02&  28 \\ 
CrII(30)    & 4824.14&          &     &          0.77&  27&    0.77&  30&    0.77&  30&    0.77&  28 \\ 
FeII(30)    & 4825.74&          &     &              &    &        &    &        &    &    0.99&  30 \\ 
FeII        & 4831.20&          &     &              &    &        &    &    0.99&  28&    0.99&  28 \\ 
CrII(30)    & 4836.24&          &     &          0.93&  22&    0.93&  31&    0.93&  31&    0.93&  29 \\ 
FeII(30)    & 4840.00&          &     &              &    &        &    &        &    &    0.99&  28 \\ 
CrII(30)    & 4848.25&          &     &          0.81&  25&    0.82&  27&    0.82&  30&    0.81&  27 \\ 
H$\beta$    & 4861.33&          &     &          0.23&  28&    0.23&  25&    0.29&  27&    0.26&  26 \\ 
FeII(25)    & 4871.27&          &     &              &    &        &    &        &  31&        &  25 \\ 
TiII(114)   & 4874.01&          &     &              &    &    0.94&  28&    0.95&  27&    0.94&  28 \\ 
CrII(30)    & 4876.40&          &     &          0.83&  25&    0.83&  31&    0.84&  32&    0.84&  30 \\ 
CrII(30)    & 4884.60&          &     &              &    &        &    &        &    &	   0.95&  30   \\ 
$[$FeII$]$4F& 4889.62&          &     &          1.02&    &        &    &    1.02&  29&    1.02&  29 \\ 
FeII(81)    & 4893.81&          &     &              &    &        &    &    0.97&  30&    0.98&  30 \\ 
CrII(190)   & 4901.65&          &     &              &    &        &    &    0.98&  27&    0.97&  30 \\ 
FeI(318)    & 4903.32&          &     &              &    &        &    &        &    &    0.98&  30 \\ 
TiII(114)   & 4911.19&          &     &          0.92&  25&    0.92&  32&    0.93&  29&    0.92&  30 \\ 
FeI(318)    & 4919.00&          &     &          0.98&  20&        &    &    0.97&  28&    0.98&  29 \\ 
FeI(318)    & 4920.50&          &     &              &    &        &    &    0.95&  31&        &     \\ 
HeI(48)     & 4921.93&          &     &              &    &        &    &    0.98&  31&    0.98&  29 \\ 
FeII(42)    & 4923.92&          &     &          0.39&  31&    0.32&  30&    0.34&  30&    0.37&  27 \\ 
BaII(1)     & 4934.08&          &     &          0.98&  24&    0.97&  31&    0.97&  29&    0.97&  31 \\ 
FeII        & 4951.59&          &     &              &    &        &    &    0.98&  30&        &     \\ 
FeI (318)   & 4957.60&          &     &          0.93&  23&    0.94&  25&    0.93&  24&    0.93&  27 \\ 
FeII        & 4977.03&          &     &              &    &    0.98&  33&        &    &        &     \\ 
FeII        & 4984.50&          &     &              &    &        &    &    0.98&  32&        &     \\ 
FeII(36)    & 4993.35&          &     &          0.94&  22&    0.93&  26&    0.93&  29&    0.94&  26 \\ 
FeII        & 5001.92&          &     &          0.92&  20&    0.92&  29&    0.92&  28&    0.93&  26 \\ 
FeII        & 5004.20&          &     &              &    &        &    &    0.97&  26&        &     \\ 
FeII        & 5007.45&          &     &              &    &        &    &        &    &    0.98&  27 \\ 
TiII(113)   & 5010.21&          &     &              &    &        &    &        &    &    0.97&  28 \\ 
TiII(71)    & 5013.69&          &     &          0.97&  26&    0.95&  27&    0.98&  28&    0.97&  30 \\ 
HeI(4)      & 5015.68&          &     &          0.97&  22&        &    &    0.98&  26&    0.97&  27 \\ 
FeII(42)    & 5018.44&          &     &          0.33&  30&    0.27&  30&    0.30&  29&    0.32&  27 \\ 
FeII        & 5026.80&          &     &              &    &        &    &        &    &    0.99&  30 \\ 
ScII(23)    & 5031.02&          &     &              &    &    0.93&  27&        &    &    0.94&  26 \\ 
FeII        & 5032.71&          &     &              &    &        &    &    0.99&  29&    0.98&  29 \\ 
FeII        & 5035.71&          &     &              &    &        &    &    0.95&  30&        &     \\ 
SiII(5)     & 5041.03&          &     &          0.89&  24&    0.88&  32&    0.88&  33&    0.89&  29 \\ 
FeII        & 5045.11&          &     &              &    &        &    &    0.99&  29&    0.99&  31 \\ 
FeII        & 5047.64&          &     &              &    &        &    &    0.98&  31&    0.98&  29 \\ 
SiII(5)     & 5056.06&          &     &          0.80&  25&    0.79&  33&    0.82&  31&        &     \\ 
FeII        & 5061.72&          &     &              &    &    0.97&  29&    0.98&  29&    0.98&  31 \\ 
FeII        & 5070.90&          &     &          0.97&  22&    0.97&  26&        &    &    0.98&  27 \\ 
TiII(113)   & 5072.30&          &     &          0.96&  23&        &    &    0.97&  28&        &     \\ 
FeII(205)   & 5074.05&          &     &          0.97&  21&	   &	&	 &    &	       &     \\ 
FeII        & 5075.77&          &     &              &    &        &    &    0.98&  32&        &     \\ 
FeII        & 5087.26&          &     &              &    &        &    &    0.97&  33&    0.99&  31 \\ 
FeII        & 5089.22&          &     &              &    &        &    &        &    &    0.99&  32 \\ 
FeII        & 5100.74&          &     &          0.90&  24&    0.91&  31&    0.90&  31&    0.91&  29 \\ 
FeII        & 5112.99&          &     &              &    &        &    &        &    &    0.99&  28 \\ 
FeII(35)    & 5120.34&          &     &          0.98&  27&    0.97&  33&    0.98&  31&        &     \\ 
TiII(86)    & 5129.16&          &     &          0.92&  25&    0.92&  30&    0.93&  29&    0.93&  29 \\ 
FeII(35)    & 5132.67&          &     &              &    &    0.96&  30&    0.98&  32&    0.98&  27 \\ 
FeII(35)    & 5136.80&          &     &          0.98&  28&    0.97&  32&    0.97&  32&    0.98&  33 \\ 
FeI(318)    & 5139.36&          &     &              &    &    0.97&  31&    0.98&  29&    0.98&  33 \\ 
FeII        & 5144.36&          &     &              &    &    0.97&  27&    0.97&  27&        &     \\ 
FeII(35)    & 5146.12&          &     &              &    &        &    &    0.96&  27&    0.96&  26 \\ 
TiII(70)    & 5154.08&          &     &              &    &        &    &    0.93&  27&        &     \\ 
$[$FeII$]$19F&5158.78&          &     &          1.05&  27&    1.02&  26&    1.05&  29&    1.06&  26 \\ 
MgI(2)      & 5167.33&      0.90&   29&              &    &    0.94&  33&        &    &    0.94&  30 \\ 
FeII(42)    & 5169.03&      0.29&   30&          0.27&  31&    0.23&  30&    0.25&  29&    0.27&  27 \\ 
MgI(2)      & 5172.69&          &     &          0.87&  26&    0.88&  33&    0.88&  33&    0.86&  32 \\ 
FeII        & 5177.39&          &     &          0.97&  25&        &    &    0.98&  29&    0.98&  27 \\ 
FeII        & 5180.32&          &     &              &    &        &    &        &    &    0.98&  30 \\ 
MgI(2)      & 5183.61&      0.84&   27&          0.85&  25&    0.86&  29&    0.86&  30&    0.86&  30 \\ 
TiII(86)    & 5185.91&      0.90&   25&          0.93&  23&    0.92&  28&    0.93&  27&    0.92&  29 \\ 
TiII(70)    & 5188.69&      0.89&   28&          0.88&  26&    0.89&  30&    0.89&  30&    0.89&  29 \\ 
FeII(49)    & 5197.58&      0.69&   30&          0.78&  29&    0.73&  30&    0.74&  30&    0.74&  26 \\ 
CrI(7)      & 5208.43&          &     &              &    &        &    &    0.97&  30&    0.98&  32 \\ 
FeII(49)    & 5234.62&      0.73&   30&          0.74&  30&    0.69&  30&    0.72&  30&    0.72&  27 \\ 
CrII(43)    & 5237.32&      0.86&   30&              &    &    0.84&  30&    0.85&  31&    0.85&  29 \\ 
ScII(26)    & 5239.82&          &     &          0.97&  24&    0.98&  28&    0.97&  29&    0.97&  30 \\ 
CrII(38)    & 5243.50&          &     &              &    &        &    &        &    &    0.98&  26 \\ 
CrII(23)    & 5249.43&          &     &              &    &    0.98&  29&        &    &    0.98&  30 \\ 
FeII(49)    & 5254.93&      0.93&   28&          0.92&  27&    0.91&  30&    0.92&  30&    0.92&  29 \\ 
FeII        & 5260.26&      0.94&   27&          0.93&  23&        &    &    0.92&  32&    0.93&  28 \\ 
FeII(48)    & 5264.80&      0.88&   26&              &    &        &    &    0.87&  27&    0.87&  25 \\ 
FeII(49)    & 5276.00&      0.70&   30&          0.72&  30&        &    &    0.71&  30&    0.72&  26 \\ 
CrII(43)    & 5279.95&          &     &              &    &        &    &    0.94&  30&    0.94&  28 \\ 
FeII(41)    & 5284.10&      0.90&   29&          0.90&  27&    0.90&  31&    0.89&  30&    0.90&  29 \\ 
FeII        & 5291.67&      0.97&   29&          0.97&  24&    0.97&  29&    0.96&  30&    0.97&  27 \\ 
CrII(24)    & 5305.86&          &     &              &    &        &    &    0.94&  29&        &     \\  
CrII(43)    & 5308.42&      0.95&   30&          0.95&  24&    0.94&  31&    0.95&  31&    0.95&  30 \\ 
CrII(43)    & 5310.69&          &     &          0.97&  22&        &    &    0.96&  30&    0.98&  28 \\                                           
CrII(43)    & 5313.58&      0.93&   29&          0.92&  24&    0.90&  30&    0.90&  30&    0.92&  28 \\ 
FeII(49)    & 5316.66&      0.62&   29&          0.63&  28&    0.60&  28&    0.59&  28&    0.60&  24 \\ 
FeII(49)    & 5325.56&      0.90&   29&          0.91&  24&    0.91&  30&    0.91&  31&    0.90&  27 \\ 
$[$FeII$]$19F&5333.65&      1.03&   26&          1.02&  26&    1.03&  22&    1.04&  27&    1.01&  24 \\ 
CrII(43)    & 5334.87&      0.94&   31&          0.93&  24&    0.93&  31&    0.93&  29&    0.92&  27 \\ 
FeII        & 5339.59&          &     &          0.96&  26&    0.96&  32&    0.97&  32&    0.96&  31 \\ 
FeII(48)    & 5362.87&      0.84&   30&          0.83&  27&    0.82&  30&    0.82&  30&    0.82&  26 \\ 
FeI(15)     & 5371.49&          &     &              &    &    1.03&  30&    1.02&  31&    1.04&  31 \\ 
TiII(69)    & 5381.03&      1.04&   27&          1.03&  24&    1.03&  27&    0.97&  25&    1.03&  26 \\ 
FeI(1146)   & 5383.38&      0.99&   31&              &    &        &    &    0.98&  25&    0.98&  29 \\ 
FeII        & 5387.07&      0.97&   32&          0.97&  24&        &    &    0.98&  31&    0.97&  28 \\ 
FeII        & 5395.86&      0.98&   28&              &    &        &    &    0.98&  29&    0.98&  29 \\ 
CrII(23)    & 5407.62&          &     &              &    &        &    &    0.98&  31&    0.97&  30 \\ 
FeII(48)    & 5414.07&          &     &              &    &    0.93&  30&    0.95&  32&        &     \\                               
TiII(69)    & 5418.78&      0.99&   27&              &    &        &    &    0.97&  30&        &     \\ 
CrII(23)    & 5420.93&          &     &              &    &        &    &    0.97&  31&    0.97&  29 \\                              
FeII(49)    & 5425.25&      0.93&   28&          0.93&  25&    0.92&  28&    0.92&  29&    0.93&  26 \\ 
FeII        & 5427.82&          &     &              &    &        &    &    0.97&  30&    0.98&  29 \\                                           
FeII(55)    & 5432.98&      0.95&   27&              &    &    0.95&  28&    0.95&  26&    0.95&  25 \\ 
FeII        & 5439.71&          &     &              &    &        &    &        &    &    0.99&  27 \\ 
FeI(15)     & 5446.91&          &     &              &    &    1.02&  33&    1.03&  31&    1.03&  30 \\ 
FeI(15)     & 5455.61&      1.03&   29&          1.02&  26&    1.03&  27&    1.02&  27&    1.03&  29 \\                                           
CrII(50)    & 5478.37&      0.94&   30&          0.95&  23&    0.95&  31&    0.95&  30&    0.95&  28 \\ 
FeII        & 5482.31&      0.97&   24&              &    &        &    &    0.96&  29&        &     \\ 
FeII        & 5487.63&      0.97&   27&              &    &        &    &    0.97&  30&    0.98&  28 \\ 
CrII(50)    & 5508.62&      0.96&   27&              &    &        &    &    0.96&  27&    0.97&  25 \\ 
CrII(23)    & 5510.71&          &     &              &    &        &    &        &    &    0.97&  29 \\ 
ScII(31)    & 5526.81&      1.05&   28&              &    &    1.04&  27&    1.03&  30&    1.05&  29 \\ 
FeII(55)    & 5534.84&      0.88&   29&          0.87&  28&    0.87&  30&    0.86&  31&    0.87&  27 \\ 
FeII(166)   & 5544.76&          &     &              &    &        &    &        &    &    0.97&  29 \\ 
FeI(686)    & 5615.64&          &     &              &    &        &    &    0.98&  31&        &     \\ 
CrII(189)   & 5620.63&          &     &              &    &        &    &        &    &    0.99&  30 \\ 
FeII(57)    & 5627.49&          &     &          0.99&  22&    0.99&  28&    0.98&  29&    0.99&  30 \\  
FeII(57)    & 5657.92&          &     &          0.97&  26&    0.98&  32&    0.97&  30&    0.97&  33 \\ 
ScII(29)    & 5684.19&          &     &              &    &        &    &        &    &    1.02&  27 \\ 
FeII(164)   & 5747.90&          &     &          0.99&  22&        &    &        &    &    0.99&  26 \\ 
DIB         & 5780.37&      0.95&   27&          0.93&  26&    0.93&  26&    0.94&  31&    0.94&  27 \\ 
DIB         & 5793.22&          &     &          0.99&  23&	   &	&	 &    &	       &     \\ 
FeII(211)   & 5795.87&          &     &          0.99&  22&	   &	&	 &    &	       &     \\ 
DIB         & 5796.96&      0.97&   38&          0.96&  35&    0.97&  33&    0.97&  34&    0.97&  36 \\ 
FeII(163)   & 5813.67&      0.97&   30&          0.98&  20&        &    &    0.98&  25&    0.98&  25 \\ 
FeII(164)   & 5823.15&          &     &          0.99&  23&    0.99&  26&        &    &    0.99&  26 \\ 
FeII(182)   & 5835.49&          &     &              &    &    0.99&  33&        &    &    0.99&  28 \\ 
DIB         & 5849.80&      0.98&   35&          0.99&  31&        &    &    0.99&  31&    0.98&  30 \\                             
HeI (11)    & 5875.72&      0.96&   30&          0.96&  24&    0.96&  26&    0.95&  30&    0.96&  31 \\ 
NaI (1)     & 5889.95&      0.06&   33&          0.03&  33&    0.04&  33&    0.07&  33&    0.08&  33 \\ 
NaI (1)     & 5895.92&      0.10&   33&          0.07&  34&    0.08&  33&    0.13&  34&    0.14&  33 \\ 
SiII(4)     & 5978.93&      0.94&   28&              &    &    0.93&  31&    0.93&  32&	       &     \\ 
FeII(46)    & 5991.37&      1.07&   26&          1.06&  26&    1.07&  28&    1.07&  29&	       &     \\ 
CrII(105)   & 6053.46&          &     &          0.99&  22&        &    & 	 &    &	       &     \\ 
FeII        & 6060.99&          &     &          0.99&  19&        &    &        &    &	       &     \\ 
FeII(46)    & 6084.10&      1.03&   26&          1.03&  25&    1.04&  26&	 &    &	       &     \\ 
NiI (45)    & 6108.12&      1.01&   28&              &	  &	   &	&	 &    &	       &     \\ 
FeII(46)    & 6113.32&      0.98&   27&          0.98&  22&    0.98&  30&        &    &        &     \\  
CaI (3)     & 6122.22&          &     &          0.99&  25&	   &	&	 &    &	       &     \\ 
BaII(2)     & 6141.73&          &     &          0.99&  21&    0.99&  29&        &    &	       &     \\ 
FeII(74)    & 6147.74&      0.93&   27&          0.92&  24&    0.92&  28&	 &    &	       &     \\ 
FeII(74)    & 6149.25&      0.93&   30&          0.93&  28&    0.93&  31&	 &    &	       &     \\ 
CaI(3)      & 6162.18&          &     &          1.02&  24&	   &	&	 &    &	       &     \\ 
FeII(200)   & 6175.16&      0.98&   29&          0.97&  23&	   &	&	 &    &	       &     \\ 
FeII(163)   & 6179.39&      0.99&   28&          0.98&  22&	   &	&	 &    &	       &     \\ 
DIB         & 6195.96&      0.98&   31&          0.98&  32&    0.97&  27&	 &    &	       &     \\ 
FeI(62)     & 6219.29&      1.01&   27& 	     &	  &	   &	&	 &    &	       &     \\ 
FeII(74)    & 6238.39&      0.94&   27&          0.94&  23&    0.93&  27&	 &    &	       &     \\ 
FeII(74)    & 6247.55&      0.88&   29&          0.87&  27&    0.85&  31&	 &    &	       &     \\ 
DIB         & 6283.85&          &     &          0.92&  34&	   &	&	 &    &	       &     \\ 
$[$OI$]$1F  & 6300.30&      1.14&   26&          1.12&  28&    1.12&  29&	 &    &	       &     \\ 
FeII        & 6317.99&          &     &              &    &    0.95&  31&	 &    &	       &     \\ 
FeII(199)   & 6331.96&      0.98&   30&              &    &    0.98&  30&	 &    &	       &     \\ 
FeI(62)     & 6335.34&      1.01&   29&     	     &	  &	   &	&	 &    &	       &     \\ 
SiII(2)     & 6347.10&      0.70&   32&          0.69&  24&    0.70&  31&	 &    &	       &     \\ 
FeI(13)     & 6358.70&      1.01&   30&          1.02&  27&    1.02&  27&	 &    &	       &     \\ 
$[$OI$]$1F  & 6363.78&      1.06&   28&          1.05&  28&    1.04&  28&	 &    &	       &     \\ 
SiII(2)     & 6371.36&      0.76&   33&          0.76&  25&    0.77&  31&	 &    &	       &     \\ 
DIB         & 6375.95&          &     &          0.99&  31&	   &	&	 &    &	       &     \\ 
DIB         & 6379.29&      0.98&   30&          0.98&  30&    0.99&  27&	 &    &	       &     \\ 
FeII        & 6383.72&          &     &          0.98&  25&	   &	&	 &    &	       &     \\ 
FeI(168)    & 6393.61&          &     &          1.01&  26&   	   &	&	 &    &	       &     \\ 
FeII(74)    & 6416.93&      1.06&   26&          0.94&  24&    1.05&  26&	 &    &	       &     \\ 
FeII(40)    & 6432.68&      1.15&   24&              &    &    1.14&  24&	 &    &	       &     \\ 
FeII        & 6442.95&          &     &          0.98&  27&	   &	&	 &    &	       &     \\ 
FeII(199)   & 6446.41&      0.98&   25&          0.97&  26&    0.98&  31&	 &    &	       &     \\ 
FeII(74)    & 6456.38&      0.81&   30&              &    &    0.80&  31&	 &    &	       &     \\ 
FeII(199)   & 6482.20&          &     &          0.96&  28&	   &	&	 &    &	       &     \\ 
FeII(40)    & 6516.08&      1.15&   26&          1.14&  24&    1.15&  26&	 &    &	       &     \\ 
MgII(23)    & 6545.97&          &     &          0.97&  26&    0.97&  29&	 &    &	       &     \\ 
H$\alpha$   & 6562.81&      1.05&   26&          0.95&  22&    1.05&  26&        &    &        &     \\ 
ScII(19)    & 6604.59&          &     &              &    &    1.02&  28&	 &    &	       &     \\ 
DIB         & 6613.56&      0.96&   33&          0.96&  30&    0.96&  31&    	 &    &	       &     \\ 
FeII        & 6627.24&      0.99&   30& 	     &	  &	   &	&	 &    &	       &     \\ 
CaI(32)     & 6717.69&          &     &              &    &    1.01&  31&	 &    &	       &     \\ 
$[$FeII$]$14F&7155.16&          &     &          1.08&  23&    1.08&  25&	 &    &	       &     \\ 
$[$CaII$]$1F& 7291.46&          &     &          1.47&  28&    1.34&  27& 	 &    &	       &     \\  
$[$CaII$]$1F& 7323.88&          &     &          1.36&  28&    1.36&  29&	 &    &	       &     \\ 
NI(3)       & 7423.64&          &     &          0.95&  25&    0.95&  34&	 &    &	       &     \\ 
NI(3)       & 7442.39&          &     &          0.92&  22&    0.93&  29&	 &    &	       &     \\ 
FeII(73)    & 7449.34&          &     &          1.03&  23&    1.03&  26&	 &    &	       &     \\ 
$[$FeII$]$14F&7452.54&          &     &          1.03&  29&    1.03&  28&	 &    &	       &     \\ 
FeII(73)    & 7462.39&          &     &          1.07&  25&    1.08&  28&        &    &        &     \\ 
NI (3)      & 7468.31&          &     &          0.89&  26&    0.90&  31&	 &    &	       &     \\ 
FeII(72)    & 7479.70&          &     &              &    &    0.99&  32&	 &    &	       &     \\ 
KI(1)       & 7664.87&          &     &          0.77&  29&    0.76&  29&	 &    &	       &     \\ 
KI (1)      & 7698.97&          &     &          0.90&  28&	   &	&	 &    &	       &     \\ 
            &        &          &     &          0.95&  37&        &    &	 &    &	       &     \\ 
OI(1)       & 7771.94&          &     &          0.68&  25&    0.67&  31&   	 &    &	       &     \\ 
MgII(8)     & 7877.05&          &     &          0.91&  26&	   &	&	 &    &	       &     \\ 
MgII(8)     & 7896.37&          &     &          0.88&  24&	   &	&	 &    &	       &     \\ 
P23         & 8345.55&          &     &          0.88&  24&    0.87&  32&	 &    &	       &     \\ 
P22         & 8359.00&          &     &          0.85&  27&    0.83&  30&	 &    &	       &     \\ 
P20         & 8392.40&          &     &          0.76&  25&    0.75&  33&  	 &    &	       &     \\ 
P19         & 8413.32&          &     &          0.71&  26&    0.71&  34&	 &    &	       &     \\ 
P18         & 8437.96&          &     &          0.68&  24&    	   &	&	 &    &	       &     \\ 
CaII(2)     & 8498.04&          &     &          1.79& 21:&	   &	&	 &    &	       &     \\  
CaII(2)     & 8542.11&          &     &          1.95& 20:&	   &	&	 &    &	       &     \\ 
NI(8)       & 8567.74&          &     &          0.94&  30&	   &	&	 &    &	       &     \\ 
P14         & 8598.39&          &     &          0.60&  25&    0.60&  30&	 &    &	       &     \\ 
NI(8)       & 8629.24&          &     &          0.90&  24&	   &	&	 &    &	       &     \\ 
CaII(2)     & 8662.17&          &     &          1.88& 23:&	   &	&	 &    &	       &     \\ 
NI(1)       & 8680.28&          &     &          0.76&  25&	   &	&	 &    &	       &     \\ 
NI(1)       & 8683.40&          &     &          0.82&  24& 	   &	&	 &    &	       &     \\ 
NI(1)       & 8686.15&          &     &          0.87&  25&	   &	&	 &    &	       &     \\ 
P12         & 8750.47&          &     &          0.60&  25&	   &	&	 &    &	       &     \\ 
MgI(7)      & 8806.76&          &     &          1.03&  26&	   &	&	 &    &	       &     \\ 
P11         & 8862.78&          &     &          0.63&  25&	   &	&	 &    &	       &     \\ 
\end{supertabular}									    			     
\label{Lines}

%\end{center}

\end{document}